\documentclass[10pt]{article}
\usepackage{amsmath}
\usepackage{graphicx}
\usepackage{amsfonts}
\usepackage{amssymb}
\usepackage{epsf}
\usepackage{latexsym}

\def\80{\hspace{0.8in}}

\def\brho{\mbox{\boldmath$\rho$}}
\def\bPi{\mbox{\boldmath$\Pi$}}

\newcommand{\be}{\begin{enumerate}}
\newcommand{\ee}{\end{enumerate}}
\newcommand{\bi}{\begin{itemize}}
\newcommand{\ei}{\end{itemize}}
\newcommand{\bd}{\begin{description}}
\newcommand{\ed}{\end{description}}

\def\beq{\begin{equation}}
\def\eeq{\end{equation}}
\def\bea{\begin{eqnarray}}
\def\eea{\end{eqnarray}}
\def\hat{\widehat}

\def\pa{\partial}
\def\d{\textrm{d}}

\def\ttA{\mbox{\tt A}}
\def\ttB{\mbox{\tt B}}
\def\ttC{\mbox{\tt C}}
\def\ttH{\mbox{\tt H}}
\def\ttL{\mbox{\tt L}}
\def\ttD{\mbox{\tt D}}
\def\ttE{\mbox{\tt E}}
\def\ttF{\mbox{\tt F}}

\def\NSI{Na\"{\i}ve Schr\"{o}dinger Interpretation }

\def\ma{\mbox{a}}

\def\md{\mbox{d}} 
\def\me{\mbox{e}}

\def\mi{\mbox{i}}

\def\ml{\mbox{l}}

\def\mn{\mbox{n}}

\def\mp{\mbox{p}} 
\def\np{\mbox{p}}

\def\muu{\mbox{u}}

\def\mD{\mbox{D}}

\def\mI{\mbox{I}}

\def\mK{\mbox{K}}
\def\mL{\mbox{L}}
\def\mM{\mbox{M}}
 
\def\mO{\mbox{O}}
\def\mP{\mbox{P}}

\def\mT{\mbox{T}}

\def\mZ{\mbox{Z}} 
\def\sa{\mbox{\scriptsize a}}

\def\sd{\mbox{\scriptsize d}}
\def\se{\mbox{\scriptsize e}}

\def\sll{\mbox{\scriptsize l}}  
\def\sm{\mbox{\scriptsize m}}
 
\def\so{\mbox{\scriptsize o}} 
\def\sp{\mbox{\scriptsize p}}
\def\sq{\mbox{\scriptsize q}}
\def\sr{\mbox{\scriptsize r}}

\def\st{\mbox{\scriptsize t}}

\def\sS{\mbox{\scriptsize S}}
\def\sT{\mbox{\scriptsize T}}

\def\barp{\bar{p}}
\def\barq{\bar{q}}
\def\barr{\bar{r}}

\def\eph(B){\mbox{\scriptsize emergent(LMB)}}

\def\bZ{\mbox{{\bf Z}}}
\def\bQ{\mbox{\sffamily q}}

\def\fE{\mbox{\sffamily E}}

\def\fH{\mbox{\sffamily H}}

\def\fS{\mbox{\sffamily S}}
\def\fT{\mbox{\sffamily T}}

\def\fV{\mbox{\sffamily V}}
\def\fW{\mbox{\sffamily W}}

\def\sfa{\mbox{\sffamily{\scriptsize a}}}

\def\sfA{\mbox{\sffamily{\scriptsize A}}}
\def\sfB{\mbox{\sffamily{\scriptsize B}}}
\def\sfC{\mbox{\sffamily{\scriptsize C}}}

\def\sfZ{\mbox{\sffamily{\scriptsize Z}}}
\def\K{Kucha\v{r} }

\def\bn{\mbox{\bf n}}

\def\bP{\mbox{\bf P}}
\def\bQ{\mbox{\bf Q}}
\def\bS{\mbox{\bf S}}

\begin{document}

\begin{titlepage}

\begin{center}

{\LARGE\bf RELATIONAL QUADRILATERALLAND.} 

\vspace{.1in}

{\LARGE\bf ANALOGUES OF ISOSPIN AND HYPERCHARGE}

\vspace{.2in}

{\bf Edward Anderson$^1$} 

\vspace{.2in}

\large {\em $^1$ Astroparticule et Cosmologie, Universit\'{e} Paris 7 Diderot} \normalsize

\vspace{.2in}

\end{center}

\begin{abstract}

I consider the momenta and conserved quantities for $\mathbb{CP}^2$ interpreted as the space of quadrilaterals.  
This builds on seminar I and II's kinematics via making use of MacFarlane's work considering the $SU(3)$-like (and thus particle physics-like) conserved quantities that occur for $\mathbb{CP}^2$.  
I perform the additional step of further interpreting that as the configuration space of all relational quadrilaterals 
and thus an interesting toy model for whole-universe, relational and geometrodynamical-analogue physics. 
I also provide the \K observables for the quadrilateral, which is a particular resolution of the Problem of Observables.  
I study HO-like and highly symmetric potentials.
I also provide some exact solutions and qualitative behaviours for dynamics on $\mathbb{CP}^2$.  
In each case, I reinterpret the results in terms of quadrilaterals.  
This paves the way for the quantum mechanical study of the relational quadrilateral and for investigations of a number of Problem of Time strategies and of a number of other foundational and qualitative investigations of Quantum Cosmology.  

\end{abstract}

\vspace{0.2in}

\noindent Seminar III on relational quadrilaterals.


\vspace{0.2in}  

\mbox{ }

\noindent $^1$ edward.anderson@apc.univ-paris7.fr

\end{titlepage}

%===========================================================================================================================================
\section{Introduction}
%===========================================================================================================================================

Paper I \cite{QuadI} provided and interpreted shape coordinates and other coordinate systems for the relational particle mechanics (RPM) \cite{RPMConcat, B03, 06II, TriCl, FORD, 08I, AF, +tri, Cones, ScaleQM, 08II, 08III, SemiclIII, FileR, AHall} up to quadrilateralland, whose shape space (relative configuration space) is $\mathbb{CP}^2$.  
See also Paper I for more details of what RPM's are for \underline{Keys 1 to 8} for unlocking RPM's up to quadrilateralland and for other applications of the $\mathbb{CP}^2$ geometry that quadrilateralland possesses.

In the present seminar, in Sec 2, I provide and interpret the shape momenta conjugate to the shape coordinates for  the various RPM theories up to quadrilateralland.
In Sec 3, I consider the corresponding Hamiltonians.  
Secs 4 and 5 consider \underline{Key 9}: the isometries for RPM's up to quadrilateralland; 
this builds on earlier work by Smith \cite{Smith60}, Franzen and I \cite{AF}.
I already studied \cite{08I, +tri} how triangleland has $SO(3) = SU(2)/\mathbb{Z}_2$ as its isometry group.  
On the other hand, $N$-stop metroland has $SO(N - 1)$ as its isometry group, among which 4-stop metroland's is SO(3) again.  
This led to spherical polar mathematics and various further analogies \cite{AF, +tri} with Molecular Physics (rigid rotors, the Stark Effect, Pauling's study of the spectra of crystals and the theory underlying Raman spectroscopy).  
\noindent In the present seminar, I consider the conserved quantities for quadrilateralland, i.e. the quadrilateralland interpretation of 
$\mathbb{CP}^2$'s isometry group, $SU(3)/\mathbb{Z}_3$ \cite{MacFarlane}.  
The quadrilateralland interpretation of $\mathbb{CP}^2$'s `$SU(3)$' of conserved quantities is one of the principal results of the present seminar.
Whilst this is no longer analogous to Molecular Physics, it is now analogous to Particle Physics.
I give hypercharge and isospin \cite{PS} analogues in terms of Gibbons--Pope-type coordinates \cite{GiPo, QuadI} and interpret these in terms of 
quadrilateralland quantities.

In \cite{RWR, FileR}, RPM's are argued in detail to be good models of geometrodynamics; thus they are useful in a number of ways for the study of the Problem of Time and for various other foundational and qualitative issues in Quantum Cosmology (extending e.g. \cite{08III, SemiclIII, FileR, QuadI}).   
One of these issues is the Problem of Observables; the present seminar resolves this in the sense of \K \cite{Kuchar93} for the quadrilateral, 
based on Papers I and II's treatment of shape coordinates and shape momenta (Sec 6). 
More Problem of Time \cite{K92, I93, APOT, FileR} applications are in Papers III \cite{QuadIII} (finishing the \NSI approach started in Paper I 
after obtaining wavefunctions) and, especially, IV \cite{QuadIV} (Histories, Semiclassical, Records and Halliwell-type combined approaches).
N.B. that the classical dynamics and QM are required prior to the Quantum Gravity and Quantum Cosmology applications; this is the main  
purpose of Papers II and III.  
In Sec 7, I give the classical equations of motion for quadrilateralland.  
In Sec 8, I consider HO-type potentials for quadrilateralland (\underline{Key 10}).  
In Sec 9, I interpret the geodesics on $\mathbb{CP}^2$ (\underline{Key 11}) in quadrilateralland terms (i.e. as a sequence of quadrilaterals).  
In Sec 10 I consider HO dynamics on $\mathbb{CP}^2$ from a qualitative perspective (\underline{Key 12}) in quadrilateralland terms.  
These free problem and HO problem cases are then considered at the quantum level in Paper III (see the Conclusion for an outline).

%=====================================================================================================================
%=====================================================================================================================
\section{Physical interpretation of the shape momenta}
%=====================================================================================================================
%=====================================================================================================================

I use relative angular momenta and relative distance momenta as names for the conjugates of relative angles and of ratios of relative separations 
respectively.  
I choose to use the first option in each case.

%=====================================================================================================================
\subsection{3- and 4-stop metroland}
%=====================================================================================================================

For 3-stop metroland in polar coordinates, the momenta are [dropping (a) labels and recycling the notation $\mp_i$ to mean the conjugate of 
$\mn^i$],
\beq
{\cal D} := p_{\varphi} = \varphi^* = \mn_1\mp_2 - \mn_2\mp_1 = {\ttD}_2\mn_1/\mn_2 - {\ttD}_1\mn_2/\mn_1 
\eeq
for $\ttD_i$ the {\it partial dilations} (in parallel to the $\mI_i$ being partial moments of inertia) $\mn^i\np_i$ (no sum).  
The second form of this is manifestly a shape-weighted {\it relative dilational quantity} corresponding to a particular exchange of dilational 
momentum between the \{bc\} and \{a\} clusters.
It is indeed conceptually clear that the conjugate to the non-angular length ratio $\varphi^{(\sfa)}$ will be a relative distance momentum.
I generally use the notation ${\cal D}$ for whichever type of relative distance momenta.  

\mbox{ }  

For 4-stop metroland in spherical coordinates, the momenta are [dropping (Hb) or (Ka) labels] 
\beq
{\cal D}_{\phi}   := p_{\phi} = \phi^* = \mn_1\mp_2 - \mn_2\mp_1  = {\ttD}_2\mn_1/\mn_2 - {\ttD}_1\mn_2/\mn_1 \mbox{ } ,   
\label{WillBeA3}
\eeq
i.e. a a weighted relative dilational quantity corresponding to a particular exchange of dilational momentum between the \{ab\} and \{cd\} 
clusters in the H-case or the \{bc\} and \{Ta\} clusters in the K-case, and
\beq
{\cal D}_{\theta} := p_{\theta} \mbox{ } . 
\eeq

%=====================================================================================================================
\subsection{Triangleland}
%=====================================================================================================================

The triangleland momenta in spherical coordinates and their interpretation are as follows (in terms of Dragt-type \cite{Dragt} coordinates and momenta): 

\noindent
\beq
{\cal D}_{\triangle} =: p_{\Theta} := \Theta^* = \mbox{dra}_1\Pi^{\sd\sr\sa}_2 -  \mbox{dra}_2\Pi^{\sd\sr\sa}_1  \mbox{ } , 
\eeq
\beq
{\cal J} =: p_{\Phi} := \mbox{sin}^2\Theta\,\Phi^* \mbox{ } . 
\eeq
Here, and more generally, I use ${\cal J}$ to denote angular momenta.  
This ${\cal J}$, moreover, clearly cannot be an overall angular momentum since $\ttL = 0$ applies.  
It is indeed a relative angular momentum \cite{08I}: 
\beq
{\cal J} = I_1I_2\Phi^{*}/I = I_1I_2\{\theta_2^* - \theta_1^*\}/I = \{I_1\mL_2 - I_2\mL_1\}/I = 
\mL_2 = - \mL_1 = \{\mL_2 - \mL_1\}/2 \mbox{ } 
\eeq
[the fourth equality uses the zero angular momentum constraint (I.8)].   
Thus this can be interpreted as the angular momentum of one of the two constituent subsystems, minus the 
angular momentum of the other, or half of the difference between the two subsystems' angular momenta.  
That is indeed a relative angular momentum also ought to also be clear from it being the conjugate of a relative angle.
The $\Theta$ and $\Phi$  coordinates represent a clean split into pure non-angle ratios and pure angle ratios, 
by which they produce one relative dilational momentum and one relative angular momentum as their conjugates.

Franzen and I \cite{AF} termed the collective set of quantities of this nature relative {\it rational momenta} since they correspond 
to the general-ratio generalization of angle-ratio's angular momenta.
Franzen and I already noted that the rational momentum concept also naturally extends to include mixed dilational momentum 
and angular momentum objects in addition to the above examples of purely dilational and purely angular objects.
Rational momenta was previously called {\it generalized angular momenta} by Smith \cite{Smith60}.  
Serna and I do not use this name since it is not conceptually descriptive; we rather unravel exactly what it means 
physically and thus call it by its `true name' \cite{WheelerInt, Kvothe}.
The rather conceptually-cleaner introduction of this at the level of the momenta rather than \cite{AF}'s at the level of the conserved quantities 
is new to the present seminar).
Our final proposal is to call them {\it shape momenta}, since what are mathematically ratio variables can also be seen to be 
dimensionless shape variables, and the quantity in question is the momentum conjugate to such a quantity.  
We celebrate this by passing from the notation ${\cal R}$ for `rational' to ${\cal S}$ for `shape'.

This `true naming' becomes clear in moving, away from the previous idea of interpreting in physical space the $SO(n)$ mathematics of 
the first few RPM models studied, to the following line of thought.  

\noindent 1) Scale--shape splits are well defined.  
Then there are corresponding splits into scale momenta and shape momenta.  

\noindent 2) The shape momenta are conjugate to dimensionless variables, i.e. ratios (or functions of ratios), accounting for why the 
previously encountered objects were termed rational momenta. 

\noindent 3) Then in some cases, shape momentum mathematics coincides with (arbitrary-dimensional) angular momentum mathematics, and also 
some shapes/ratios happen to be physically angles in space, so the interpretation in space {\sl indeed is} as angular momentum.  
  
\noindent 4) But in other cases, shapes can correspond physically to ratios other than those that go into angles in space, 
e.g. ratios of two lengths (then one's momentum is a pure relative distance momentum) or a mixture of angle and 
non-angle in space ratios (in which case one has a general shape momentum).
Moreover, there is no a priori association between shape momenta and $SO(n)$ groups; this {\sl happens} to be the case for the first few examples encountered ($N$-stop metroland, triangleland) but ceases to be the situation for quadrilateralland (and $N$-a-gonlands beyond that).

%=====================================================================================================================
\subsection{Quadrilateralland}
%=====================================================================================================================

The Gibbons--Pope-type coordinates for quadrilateralland extend the above triangleland spherical polar coordinates 
in constituting a clean split into pure non-angle ratios and pure angle ratios (two of each).
Thus their conjugates are again cleanly-split pure relative angular momenta and relative dilational momenta as their conjugates (two of each): 
\beq
{\cal J}_{\psi}  =: p_{\psi}  = \mbox{sin}^2\chi\mbox{cos}^2\chi\{\psi^* + \mbox{cos}\,\beta\,\phi^*\}/4 \mbox{ } , \hspace{1.15in}
\eeq
\beq
{\cal J}_{\phi}  =: p_{\phi}  = \mbox{sin}^2\chi\{\mbox{cos}^2\chi\{\phi^* + \mbox{cos}\,\beta\,\psi^*\} + \mbox{sin}^2\chi\,\mbox{sin}^2\beta\, \phi^*\}/4 \mbox{ } , \hspace{0.1in} 
\eeq
\beq
{\cal D}_{\beta} =: p_{\beta}  =  \mbox{sin}^2\chi\,\beta^*/4 \mbox{ } , \hspace{2.25in}
\eeq
\beq
{\cal D}_{\chi}  =: p_{\chi}  = \chi^* \mbox{ } . \hspace{2.75in} 
\eeq
\mbox{ } \mbox{ } The interpretation of these in terms of quadrilaterals are as follows (using Fig I.1's and Sec I.7's nomenclatures).  
\noindent The basic H = H(DD)'s $p_{\beta}$ is then the relative dilation of the two posts -- universe contents, 
                     whilst its $p_{\chi}$ is the relative dilation of the posts contents relative to their `universe separation'.
\noindent H(M$^*$D)'s $p_{\beta}$ is then the relative dilation of the selected post to the `universe separation' crossbar, 
           whilst its $\chi$ is the relative dilation of the selected-post-and-crossbar to the non-selected post. 
\noindent The basic K = K(T)'s $p_{\beta}$ is then the relative dilation of the back to the seat (ie a change of sharpness/flatness shape momentum for the obvious triangle subsystem in $d > 1$), whilst $p_{\chi}$ is the relative dilation of the back-and-seat (for the obvious triangle subsystem) to the remaining leg particle - i.e. a relative dilation of the whole triangle relative to the separation between it and the remaining particle.
\noindent K(M$^*$D)'s $p_{\beta}$ is then the relative dilation of the seat to the leg, whilst its $p_{\chi}$ is the relative dilation of the seat-and-leg to the back binary.   
\noindent Finally, K(M$^{\sT}$D)'s $p_{\beta}$ is then the relative dilation of the back to the leg, whilst its $p_{\chi}$ is the relative 
dilation of the back-and-leg with respect to the seat.   
\noindent 
In each case, $p_{\phi}$ and $p_{\psi}$ are a co-rotation and a counter-rotation of the two selected objects without any discernible pattern.

%=======================================================================================================================================
%=======================================================================================================================================
\section{Forms of the shape Hamiltonians}
%=======================================================================================================================================
%=======================================================================================================================================

The corresponding Hamiltonians are 
\beq
\fH = p_{\varphi}\mbox{}^2/2 + \fV = {\cal D}^2/2 + \fV \mbox{ (3-stop metroland) ,}
\eeq
\beq
\fH = p_{\theta}\mbox{}^2/2 + p_{\phi}\mbox{}^2/2\,\mbox{sin}^2\theta + \fV = {\cal D}_{\theta}\mbox{}^2/2 + {\cal D}_{\phi}\mbox{}^2/2\,\mbox{sin}^2\theta + \fV \mbox{ } \mbox{ (4-stop metroland) } , 
\eeq
\beq
\fH = p_{\Theta}\mbox{}^2/2             +   p_{\Phi}\mbox{}^2/2\,\mbox{sin}^2\Theta + \fV
     = {\cal D}_{\triangle}\mbox{}^2/2   +   {\cal J}^2/2\,\mbox{sin}^2\Theta + \fV \mbox{ } \mbox{ (triangleland) and } 
\eeq
$$
\fH = \frac{p_{\chi}^2}{2} + \frac{2}{\mbox{sin}^2\chi}\left\{p_{\beta}^2 + \frac{1}{\mbox{sin}^2\beta}
\{p_{\phi}^2  \mbox{ } +   \mbox{ } p_{\psi}^2  \mbox{ } -  \mbox{ } 2\, p_{\phi}p_{\psi} \mbox{cos}\,\beta \}\right\} + \frac{2}{\mbox{cos}^2\chi}p_{\psi}^2  + \fV \hspace{0.4in}
$$
\beq
=\frac{{\cal D}_{\chi}^2}{2} + \frac{2}{\mbox{sin}^2\chi}\left\{{\cal D}_{\beta}^2 + \frac{1}{\mbox{sin}^2\beta}
\{{\cal J}_{\phi}^2 + {\cal J}_{\psi}^2 - 2\,{\cal J}_{\phi}{\cal J}_{\psi}\mbox{cos}\,\beta \}\right\} + \frac{2}{\mbox{cos}^2\chi}
{\cal J}_{\psi}^2  + \fV \mbox{ } \mbox{ (quadrilateralland) } . 
\eeq

%========================================================================================================
%========================================================================================================
\section{Physical interpretation of RPM's relationalspace isometries/conserved quantities}\label{SSec: Cons}
%========================================================================================================
%========================================================================================================

\noindent For a dynamical system, conserved quantities correspond to isometries of the kinetic metric that are also 
respected by the potential.
This Section and the next deal with isometries; the subdivision of the generators of these into conserved quantities 
and elsewise for various potentials is the subject of Sec 7.  
\beq
\mbox{Isom($\fS(N, 1))$ = Isom($\mathbb{S}^{n - 1}$) = $PSO(n)$ (the $n$-dimensional projective special orthogonal group) = $SO(n)$ } .
\eeq
\beq
\mbox{Isom($\fS(N, 2))$ = Isom($\mathbb{CP}^{n - 1}$) = $PSU(n)$ (the $n$-dimensional projective special unitary group) = $SU(n)/\mathbb{Z}_n$ } . 
\eeq
The $SU(n)$ versus  $SU(n)/\mathbb{Z}_n$ distinction does not affect the algebra involved, though it does matter as regards 
some further subtleties (along the lines of the much better-known $SU(2)$ versus $SU(2)/\mathbb{Z}_2 = SO(3)$ distinction).  
The above is standard to fairly standard mathematics; moreover, the {\sl physical interpretation} of the generators of these 
(which for various classes of potentials are to be interpreted as conserved quantities) is somewhat unusual, as I shall 
build up case-by-case below.

%=====================================================================================================================
\subsection{$N$-stop metroland cases}\label{Sprintina}
%=====================================================================================================================

The pure-shape case of 3-stop metroland is relationally trivial as per \cite{FileR}, but it is part of dynamically 
nontrivial scaled 3-stop metroland problem.  
Here the generator of Isom$(\fS(3, 1))$ = Isom$(\mathbb{S}^1) =  SO(2) = U(1)$ is just the above-described ${\cal D}\mi\ml$.   
This is mathematically the `component out of the plane' of  `angular momentum', albeit in {\sl configuration space}, there clearly being no meaningful physical concept of angular momentum in 1-$d$ space itself.

For 4-stop metroland, the three generators of Isom($\fS(4, 1)$) = Isom($\mathbb{S}^2$) = $SO(3)$ are 
\beq
{\cal D}_i = {{\epsilon_{ij}}^k}\mn^j\mp_{k} \mbox{ } .
\label{DDD}
\eeq 
(\ref{DDD}) are mathematically the three components of `angular momentum' albeit again in configuration space.  
Their physical interpretation (for the moment in the setting of H-coordinates) in space is an immediate extension of that of the already-encountered 3-component of this object (\ref{WillBeA3}):
\beq
{\cal D}_i = {\ttD}_k\mn^j/\mn^k - {\ttD}_j\mn^k/\mn^j \mbox{ } .  
\eeq
Moreover, this example's interpretation relies, somewhat innocuously, on the three conserved quantities ${\cal D}\mi\ml_i$ corresponding to three mutually perpendicular directions (the three DD axes picked out by using H-coordinates), as is brought out more clearly by the next example.

For 4-stop metroland in K-coordinates one has the above formulae again [dropping (Ka) labels instead of (Hb) ones]. 
They are clearly still all relative distance momenta, albeit corresponding to a different set of ratios.  
Then e.g. ${\cal D}_3$ is a (weighted) {\it relative dilational quantity} corresponding to a particular exchange of dilational momentum between the \{12\} and \{T3\} clusters. 
Here, one needs to use an axis system containing only one T-axis, e.g. a \{$\mT, \mM^*\mD, \mM^*\mD$\} axis system (c.f. 
Fig I.4).

For 4-stop metroland the {\it total shape momentum} counterpart of the total angular momentum is 

\noindent ${\cal D}_{\sT\so\st} = \sum_{i = 1}^{3}{\cal D}_i\mbox{}^2 = {\cal D}_{\theta}\mbox{}^2 + \mbox{sin}^{-2}\theta\,
{\cal D}_{\phi}\mbox{}^2 = 2\fT$ in terms of momenta. 
For 3-stop metroland, this is just ${\cal D}_{\sT\so\st} = {\cal D}^2$. 
In each case, finally, $\fH = {\cal D}_{\sT\so\st}/2 + \fV$.

The above pattern repeats itself, giving,  for $N$-stop metroland, $n$ -- 1 hyperspherical coordinates interpretable as a sequence of ratios of relative inter-particle cluster separations, shape space isometry group $SO(n)$ and a set of $n(n - 1)/2$ isometry generators which are, mathematically, components of `angular momentum' in configuration space.

%=====================================================================================================================
\subsection{Triangleland case}\label{Sprague}
%=====================================================================================================================

Here, the three Isom $(\mathbb{S}^2) = \mbox{Isom}\fS(3, 2) = SO(3) = SU(2)/\mathbb{Z}_2$ generators are given by 
\beq
{\cal S}_{\sfA} = {{\epsilon_{\sfA\sfB}}^{\sfC}}\mbox{dra}^{\sfB} \Pi^{\sd\sr\sa}_{\sfC} \mbox{ } ,
\label{SSS}
\eeq
which are mathematically the three components of `angular momentum' albeit yet again in configuration space rather than in space.  
Now on this occasion, there is a notion of relative angular momentum in space.
There are even three natural such, one per clustering: ${\cal J}_{(\sa)}$.    
Are these the three components of ${\cal S}_{\sfA}$?
No! 
These  three are coplanar and at 120 degrees to each other, so only can only pick one of these for any given orthogonal coordinate basis, 
much as in the above K-coordinate example. 
The other components point in an E and an S direction (c.f. Fig I.5).
E and D are then the two main useful choices of principal axes, furnishing the \{E, D, S\} and \{D, E, S\}.   
Moreover the component pointing in the D direction has the form of a  pure relative angular momenta, 
${\cal S}_3 = {\cal J}$ of the \{23\} subsystem relative to the 1 subsystem.   
The other two ${\cal S}_{\sfA}$'s are mixed dilational and angular momenta with shape-valued coefficient [dropping (a) labels]: 
\beq
\mbox{sin}\,\Phi\, {\cal D}_{\triangle} + \mbox{cos}\,\Phi\,\mbox{cot}\,\Theta\,{\cal J}   \mbox{ } \mbox{ and } 
\mbox{ } \mbox{ }
-\mbox{cos}\,\Phi\, {\cal D}_{\triangle} + \mbox{sin}\,\Phi\,\mbox{cot}\,\Theta\,{\cal J}   \mbox{ } .
\eeq 
For triangleland, the {\it total shape momentum} counterpart of the total angular momentum is 

\noindent 
${\cal S}_{\sT\so\st} = \sum_{\sfA = 1}^3{\cal S}_{\sfA}\mbox{}^2 = {\cal D}_{\triangle}\mbox{}^2 + \mbox{sin}^{-2}\Theta\,{\cal J}^2  = 2\fT$. 
Finally, $\fH = {\cal D}_{\sT\so\st}/2 + \fV$.

%=========================================================================================
%=========================================================================================
\section{Quadrilateralland case} 
%=========================================================================================
%=========================================================================================

Quadrilateralland's isometry group is Isom($\fS(4, 2)$) = Isom($\mathbb{CP}^2$) = $PSU(3)$ = $SU(3)/\mathbb{Z}_3$, 
giving the same representation theory and mathematical form of conserved quantities as in the idealized flavour 
$SU$(3) or the colour $SU$(3) of Particle Physics [these {\sl also} have this quotienting].   
MacFarlane studied this and the difference between it and $SU(3)$ in \cite{Macfarlane68}; 
they share  the same algebra, but some topological differences.
There are some parallels with the extent of the similarities between $SU(2)$ and $SU(2)/\mathbb{Z}_2 = SO(3)$ [which itself is relevant to RPM's via Isom($\fS(3, 2)$) = Isom($\mathbb{S}^1$) =  Isom($\mathbb{CP}^1$) = $PSU(2)$ = $SU(2)/\mathbb{Z}_2$].

%============================================================================================================================
\subsection{Particle Physics analogues}
%============================================================================================================================

Analogy 1) Flavour symmetry (constitution of hadrons in terms of up, down and strange quarks).\footnote{Hence going from 
%%%%%%%%%%%%%%%%%%%%%%%%%%%%%%%%%%%%%%%%%%%%%%%%%%%%%%%%%%%%%%%%%%%%%%%%%%%%%%%%%%%%%%%%%%%%%%%%%%%%%%%%%%%%%%%%%%%%%%%%%%%%%%%%
1-$d$ to 2-$d$ or (3, 2) to ($N$ $>$ 3, 2) parallels the transition of theoretical physics from the theoretical 
chemistry of Mendeleev through to the 1930's to the particle physics of the 60's through to the present day (including GUT's).}  
%%%%%%%%%%%%%%%%%%%%%%%%%%%%%%%%%%%%%%%%%%%%%%%%%%%%%%%%%%%%%%%%%%%%%%%%%%%%%%%%%%%%%%%%%%%%%%%%%%%%%%%%%%%%%%%%%%%%%%%%%%%%%%%%    
%
Our use of 1, 2, 3, +, and -- is the standard one of $SU(2)$ mathematics.
$SU(3)$ contains three overlapping such ladders (in fact three overlapping $SU(2) \times U(1)$'s, with the 
$SU(2)$'s being isospins $I_+$, $I_-$, $I_3$, $V_+$, $V_-$, $V_3$ and $U_+$, $U_-$, $U_3$ and the 
$U(1)$'s being hypercharges $Y$, $Y_V$ and $Y_U$).  
The usual set of independent such objects, $I_3$, $I_+$, $I_-$, $V_+$, $V_-$, $U_+$, $U_-$ 
and ${Y}$, are then  represented by the Gell-Mann $\lambda$-matrices up to proportion. 
Then one can obtain ${V}_3$, ${U}_3$ ${Y}_{U}$,and ${Y}_{V}$ in terms of these, these other quantities being 
useful on grounds of even-handedness between the three $SU(2) \times U(1)$'s (see the next subsection).
One can then define ${I}_{\sT\so\st } = \sum_{{\cal A} = 1}^3{I}_{{\cal A}}\mbox{}^2$, 
                    ${U}_{\sT\so\st } = \sum_{{\cal A} = 1}^3{U}_{{\cal A}}\mbox{}^2$ and 
                    ${V}_{\sT\so\st } = \sum_{{\cal A} = 1}^3{V}_{{\cal A}}\mbox{}^2$.
In total, $SU(3)$ has 3 independent commuting quantities, which are usually taken to be ${I}_{\sT\so\st }$, ${I}_3$, ${Y}$.   

\mbox{ }  

\noindent Note 1) Flavour symmetry is broken by mass differences, as it is only an approximate symmetry.  

\noindent Note 2) The word `hypercharge' leaves something to be desired via not being particularly descriptive.  
In flavour physics, it is an `extra charge' that partly contains strangeness, unlike the isospin which is pure up and down.  
Thus Serna and I prefer `strange charge' and `extra charge' as names for it (taking due note that the charge is used to imply the more 
common $U(1)$ symmetry rather than generalized non-abelian symmetry).  
In particular, we like `extra charge' due to it coming picked out alongside, but not within, the $SU(2)$ in the $SU(2) \times U(1)$ combination selected by the basis.

\mbox{ } 

\noindent Analogy 2) Colour symmetry.
This use of $SU(3)$ differs in being postulated to be exact, and in the red, green and blue labels being frivolous choices, 
so that one really has $SU(3)/\mathbb{Z}_3$.

%============================================================================================================================
\subsection{\underline{Key 9}: Quadrilateralland's conserved quantities}
%============================================================================================================================

I calligraphize all of the above symbols in the quadrilateralland case, to distinguish these quantities clearly from 
their particle physics analogues.  
This application is fact more like colour physics than approximate flavour physics, in that the symmetry is exact. 
However, whilst for uninterpreted $\mathbb{CP}^2$ one can take the three types of ladder to be frivolous labels and so involve 
$SU(3)/\mathbb{Z}_3$, the quadrilateralland interpretation pins distinction upon the three ladders, so that one wishes for the whole 
$\mathbb{CP}^2$ with its three uniform states per hemi-$\mathbb{CP}^2$ rather than a folded-up version in which the three coincide.  

%\mbox{ }  

On the basis of the above discussion, Serna and I call ${\cal I}_3$ the {\it angular charge} and ${\cal Y}$ the {\it extra angular charge}    
due to its coming alongside the usual angle charge's $SU(2)$ but not within it, as a picked out $SU(2) \times U(1)$.
In the $\mathbb{CP}^2$ realization of $SU(3)$, this is  \cite{MF79, MacFarlane} not only picked out by the basis but also by the Gibbons--Pope-type coordinates in use.  

\noindent
\beq
{\cal Y} = 2p_{\psi} = 2{\cal J}_{\psi} \mbox{ } \mbox{ } , \mbox{ }  \mbox{ } {\cal I}_3 = {p}_{\phi} = {\cal J}_{\phi} \mbox{ } .
\eeq
In terms of the quadrilateralland-significant inhomogeneous bipolar coordinates, these are then 
\beq
{\cal Y} = - 2\{ p_{\Phi_1} + p_{\Phi_2}  \} \mbox{ } , \mbox{ }  \mbox{ }
{\cal I}_3 = p_{\Phi_2} - p_{\Phi_1} \mbox{ } .  
\eeq
The meanings of ${\cal Y}$ and ${\cal I}$ are immediately inherited from those of $p_{\psi}$ and $p_{\phi}$ given in Sec 2.4.

%================================================================================================================
\subsection{Generators of the isometries in $\mZ^{\barp}$ coordinates from Noether's theorem}
%================================================================================================================

Conserved quantities in terms of $\mZ^{\barp}$ and $\Pi_{\barp}$ are presented below.  
MacFarlane \cite{MacFarlane} derived these from the Euler--Lagrange action.
The present seminar uses instead the Jacobi-type action, the outcome from which is equivalent to MacFarlane's result by the following Lemma. 

\mbox{ } 

\noindent{\bf Lemma}.  
The quantities arising from Noether's theorem as applied to a Jacobi-type action 
are equivalent to those arising from the corresponding Euler--Lagrange-type action.

\mbox{ } 

\noindent Thus MacFarlane's results carry over to the context of relational Jacobi-type actions, and provide the following conserved quantities.
\beq
2i{\cal I}_3 = \bPi \tau_3 {\bZ} - \overline{\bPi}\tau_3\overline{\bZ} \mbox{ } , \mbox{ } \mbox{ } 
 i{\cal Y} = \bPi \cdot {\bZ} - \overline{\bPi}\cdot\overline{\bZ} 
\eeq
\beq
i{\cal I}_+ = \Pi_1\mZ^2 - \overline{\Pi}_2\overline{\mZ}^1 \mbox{ } , \mbox{ } \mbox{ } 
i{\cal I}_- = \Pi_2\mZ^1 - \overline{\Pi}_1\overline{\mZ}^2 \mbox{ } ,
\eeq
\beq
i{\cal V}_+  = \Pi_1 + \overline{\bPi}\cdot\overline{\bZ}\,\overline{\mZ}^1 \mbox{ } , \mbox{ } \mbox{ } 
-i{\cal V}_- = \overline{\bPi}_1 + \bPi\cdot{\bZ}\,\mZ^1 \mbox{ } , 
\eeq
\beq
i{\cal U}_+  = \Pi_2 + \overline{\bPi}\cdot\overline{\bZ}\,\overline{\mZ}^2 \mbox{ } , \mbox{ } \mbox{ } 
-i{\cal U}_- = \overline{\bPi}_2 + \bPi\cdot{\bZ}\,\mZ^2 \mbox{ } , 
\eeq
where $\tau_3 = \mbox{\Huge(}\stackrel{\mbox{1\,\,\,\,\,\,\,0}}{0\,-1}\mbox{\Huge)}$ (the third Pauli matrix).

\mbox{ } 

\noindent Note 1) The above constitutes a nonlinear realization of the $SU(3)$. 

\noindent Note 2) For the triangleland counterpart,    
${\cal Y}_{\cal U} = -\frac{1}{4}\{{\cal I}_3 + 3{\cal Y}\}$, 
${\cal Y}_{\cal V} = -\frac{1}{4}\{{\cal I}_3 - 3{\cal Y}\}$, 
${\cal U}_3 =  \frac{1}{2}\{{\cal I}_3 - {\cal Y}\}$ and 
${\cal V}_3 =  \frac{1}{2}\{{\cal I}_3 + {\cal Y}\}$.  

\noindent The generators are of types
\beq
\Pi \mZ - \overline{\Pi}\overline{Z}          \mbox{ } , \mbox{ } \mbox{ } 
{\Pi} + \overline{\mZ}^2\overline{\Pi}        \mbox{ } , \mbox{ } \mbox{ } 
\overline{\Pi} + \mZ^2\Pi                      \mbox{ } ,
\eeq
i.e., respectively, what ${\cal I}_3$ and ${\cal Y}$, ${\cal U}_+$ and ${\cal V}_+$, and ${\cal U}_-$ and ${\cal V}_-$ pairwise collapse to; 
${\cal I}_{\pm}$ cease to exist at all.  
Quantities proportional to these generators are then a ${\cal J}_3$ and ${\cal J}_{\pm}$ for the triangle.  

\noindent Note 3) The three $SU(2)$ ladders correspond to the three triangles (or two triangles and a rhombus) of coarse-graining in Fig I.8 
[each of which, of course, is associated with a coarse-grained shape space sphere whose isometry group is the corresponding $SO(3)$.]
Furthermore, each ladder is paired with a hypercharge-type quantity to form three overlapping embedded $SU(2) \times U(1)$ 's.  
${\cal I}_1$, ${\cal I}_2$, ${\cal I}_3$, ${\cal Y}$ is one of the embedded $SU(2) \times U(1)$ groups within the $SU(3)$, the others 
being the ${\cal U}$ and ${\cal V}$ counterparts of this.

%==========================================================================================================================
\subsection{Generators of the isometries in terms of Gibbons--Pope-type momenta}
%==========================================================================================================================

\beq
\mbox{As well as} \hspace{1in}
{\cal Y} = 2p_{\psi} = 2{\cal J}_{\psi} = - 2\{ p_{\Phi_1} + p_{\Phi_2}  \} \mbox{ } \mbox{ } , \mbox{ }  \mbox{ } 
{\cal I}_3 = {p}_{\phi} = {\cal J}_{\phi} = p_{\Phi_2} - p_{\Phi_1}\mbox{ } , \hspace{3in}  
\eeq
\beq
\mbox{one has} \hspace{1in}
{\cal I}_1 = 
- \mbox{sin}\,\phi\, p_{\beta} + \frac{\mbox{cos}\,\phi}{\mbox{sin}\,\beta}\{p_{\psi} - \mbox{cos}\,\beta\,p_{\phi}\} = 
- \mbox{sin}\,\phi\, {\cal D}_{\beta} + \frac{\mbox{cos}\,\phi}{\mbox{sin}\,\beta}\{{\cal J}_{\psi} - \mbox{cos}\,\beta\,{\cal J}_{\phi}\}                                                       \mbox{ } , \mbox{ } \hspace{3in}  
\eeq
\beq
{\cal I}_2 = 
\mbox{cos}\,\phi\,p_{\beta} + \frac{\mbox{sin}\,\phi}{\mbox{sin}\,\beta}\{p_{\psi} - \mbox{cos}\,\beta\,{p}_{\phi}\} =
\mbox{cos}\,\phi\,{\cal D}_{\beta} + \frac{\mbox{sin}\,\phi}{\mbox{sin}\,\beta}\{{\cal J}_{\psi} - \mbox{cos}\,\beta\,{\cal J}_{\phi}\} \mbox{ } .  
\eeq 
\beq
\mbox{Finally, } \hspace{0.3in} 
{\cal I}_{\sT\so\st} := {\cal I}^2 = 
p_{\beta}\mbox{}^2 + \frac{1}{\mbox{sin}^2\beta}
\{p_{\phi}\mbox{}^2 - 2\mbox{cos}\,\beta\, p_{\psi}{p}_{\phi} + p_{\psi}\mbox{}^2\} = 
{\cal D}_{\beta}^2 + \frac{1}{\mbox{sin}^2\beta}\{{\cal J}_{\phi}^2 - 2\mbox{cos}\,\beta\,{\cal J}_{\phi}{\cal J}_{\psi} + {\cal J}_{\psi}^2\} 
\mbox{ } .  \hspace{2in}  
\eeq
Note 1) Thus, whether for H's or for K's there is also a pair of coordinates $\beta$ and $\chi$: additionally dependent on only one corresponding  ratio of relative separations, i.e. the ${\cal I}_1$ and ${\cal I}_2$ depend on $\beta$ alone rather than on $\chi$.  
These are conjugate to quantities that involve relative distance momenta in addition to relative angular momenta.

\noindent Note 2) The other expressions (${\cal U}_{\pm}, {\cal V}_{\pm}$) are much more complex and less insightful in these particular 
${\cal I}$-adapted Gibbons--Pope type coordinates.
Of course, ${\cal U}$ and ${\cal V}$ adapted Gibbons--Pope type coordinates exist as well, via omitting in each case a different 
choice of Jacobi vector. 
E.g. in the Jacobi H case, ${\cal I}$ is tied to the collapse to the rhombus, with ${\cal U}$ and ${\cal V}$ corresponding to the two one-post collapse triangles.
In  terms of each of these coordinate systems, the corresponding sets of picked-out $SU(2) \times U(1)$ 
quantities (i.e. \{${\cal U}_{\pm}, {\cal U}_3, {\cal Y}_{\cal U}, {\cal U}_{\sT\so\st}$\} and  
                 \{${\cal V}_{\pm}, {\cal V}_3, {\cal Y}_{\cal V}, {\cal V}_{\sT\so\st}$\} have the same expressions as above 
(with ${\cal U}$, ${\cal V}$ labels,respectively, understood but dropped on the Gibbons--Pope type coordinates in use.)

%==========================================================================================================================
\subsection{Quantum-mechanical operator expressions for these isometry generators}
%==========================================================================================================================

See \cite{AF,08I,+tri,FileR} for the forms these take in metrolands and triangleland.  
For quadrilateralland (using $\hbar = 1$), 
\beq
{\cal Y} = -2i \frac{\pa}{\pa\psi} \mbox{ } \mbox{ } , \mbox{ }  \mbox{ } \hat{\cal I}_3 = -i\frac{\pa}{\pa\phi} \mbox{ } .
\eeq
In terms of the quadrilateralland-significant inhomogeneous bipolar coordinates, these are
\beq
\hat{\cal Y} = 2i\left\{ \frac{\pa}{\pa\Phi_1} + \frac{\pa}{\pa\Phi_2}  \right\} \mbox{ } \mbox{ and } \mbox{ } \mbox{ }
\hat{\cal I}_3 = -  i \left\{\frac{\pa}{\pa\Phi_2} - \frac{\pa}{\pa\Phi_1}\right\} \mbox{ } .  
\eeq
\beq
\mbox{Also,   } \hspace{0.4in}
i\hat{\cal I}_1 = - \mbox{sin}\,\phi\frac{\pa}{\pa \beta} + \frac{\mbox{cos}\,\phi}{\mbox{sin}\,\beta}
\left\{
\frac{\pa}{\pa\psi} - \mbox{cos}\,\beta\frac{\pa}{\pa\phi}
\right\} 
\mbox{ } \mbox{ } , \mbox{ } 
i\hat{\cal I}_2 = \mbox{cos}\,\phi\frac{\pa}{\pa \beta} + \frac{\mbox{sin}\,\phi}{\mbox{sin}\,\beta}\left\{\frac{\pa}{\pa\psi} - 
\mbox{cos}\,\beta\frac{\pa}{\pa\phi}\right\} \mbox{ } .  \hspace{2in}
\eeq 
\beq
\mbox{Finally,} \hspace{0.8in}
\hat{\cal I}^2 = -
\left\{
\frac{1}{\mbox{sin}\,\beta}  \frac{\pa}{\pa\beta}\mbox{sin}\,\beta\frac{\pa}{\pa\beta} + 
\frac{1}{\mbox{sin}^2\beta}
\left\{
\frac{\pa^2}{\pa\phi^2} - 
2\mbox{cos}\,\beta\frac{\pa}{\pa\psi}\frac{\pa}{\pa\phi} + \frac{\pa^2}{\pa\psi^2}
\right\}
\right\}
\hspace{3in}  
\eeq
is also needed for the subsequent QM application \cite{MacFarlane, QuadIII}.  
I give this in the operator-ordering that is relevant for Paper III's time-independent Schr\"{o}dinger equation to be in terms of the Laplacian.
This is motivated by essentially amounting to constructing the argued-for (Sec I.1) conformal operator ordering, since the two are out by just a constant.

%===================================================================================================================================
\subsection{Interpretation: the collapse of the above to the usual $SU(2)$ operators for the ``${\cal I}$" coarse-graining}
%===================================================================================================================================

For $\rho_3 = 0$, i.e. $\chi = \pi/2$, one recovers the usual ${\cal I}_{\Gamma}$, $\Gamma = 1$ to $3$, of $SU(2)$ with 
$\beta$ playing the role of $\theta$. 
I.e. at the QM level at which these expressions are most familiar, 
\beq
{\cal I}_1 = i
\left\{
\mbox{sin} \,\Phi\, \frac{\pa}{\pa\Theta} + \mbox{cos}\,\Phi\,\mbox{cot}  \, \Theta\, \frac{\pa}{\pa\Phi}   
\right\} = i{\cal S}_2
\mbox{ } , \mbox{ } \mbox{ }
{\cal I}_2 = i
\left\{
-\mbox{cos}\,\Phi\, \frac{\pa}{\pa\Theta} + \mbox{sin}\,\Phi\,\mbox{cot}\,\Theta\, \frac{\pa}{\pa\Phi} 
\right\}
\mbox{ } , \mbox{ } \mbox{ }
{\cal I}_3 = -i\frac{\pa}{\pa\Phi} \mbox{ } .  
\eeq
[The slight disalignment is due to different axis conventions and coefficients between the inherited-from-$SU(3)$ case and 
the straight $SU(2)$ case.]  
N.B. that this case is not a triangle; it is the rhombic coarse-graining.

%=============================================================================================================================================
\subsection{Interpretation: quadrilateralland isometry generators}
%=============================================================================================================================================

I mention the parallel with ${\cal J} = -i\pa/\pa\Phi$ pure relative angular momentum in in triangleland whilst ${\cal R}_1$ and ${\cal R}_2$ 
are mixtures of relative angular momentum and relative dilational momentum.   
There is a looser parallel with ${\cal D} = -i\pa/\pa\phi$ in 4-stop metroland which has, however, a different meaning.   

\noindent
In H [=H(DD)] coordinates, the meaning of ${\cal I}_1$ and ${\cal I}_2$ coordinates is that of mixed relative angular momentum and relative  
dilation of the $\beta$ type, i.e. a rate of  change in the contents inhomogeneity (the ratio of the sizes of the two constituent subclusters).  
The meaning of ${\cal I}_{\sT\so\st}$ is the total angular momentum of the third, rhombic, coarse-graining triangle of the H in Fig I.8.  
In each case, changing which ratios one regards as primary gives similar presentations for the ${\cal U}$'s and ${\cal V}'$s.  

\noindent
In H(M$^*$D) coordinates, the meaning of ${\cal I}_1$ and ${\cal I}_2$ is that of mixed relative angular momentum and relative dilation of the 
$\beta$ type, i.e. a rate of change in the ratio of the selected post to the crossbar. 
The meaning of ${\cal I}_{\sT\so\st}$ is the total angular momentum of the first or second coarse-graining of H in Fig I.8 (depending on which post 
is selected).  

\noindent
In K [= K(T)] coordinates, the meaning of ${\cal I}_1$ and ${\cal I}_2$  is that of mixed relative angular momentum and relative dilation of the 
$\beta$ type, i.e. a rate of  change in the ratio of the back to the seat (ie a sharpness/flatness shape quantity for the obvious triangle subsystem).         
The meaning of ${\cal I}_{\sT\so\st}$ is the total angular momentum of the second coarse-graining triangle of K in Fig I.8.  

\noindent
In K(M$^*$D)-coordinates, the meaning of ${\cal I}_1$ and ${\cal I}_2$  is that of mixed relative angular momentum and relative dilation of the $\beta$ type, a rate of change of the ratio of the seat to the leg
The meaning of ${\cal I}_{\sT\so\st}$ in K(M$^{\sT}$D)  -coordinates is the total angular momentum of the third coarse-graining triangle of K 
in Fig I.8. 

\noindent
The meaning of ${\cal I}_1$ and ${\cal I}_2$ in H(M$^*$D) coordinates is that of mixed relative angular momentum and relative dilation of the $\beta$ type, i.e. a rate of change in the ratio of the back to the leg.  
The meaning of ${\cal I}_{\sT\so\st}$ in H-coordinates is the total angular momentum of the first coarse-graining triangle of K in Fig I.8.  
In each case, changing which ratios one regards as primary gives similar presentations for the ${\cal U}$'s and ${\cal V}'$s.

%===================================================================================================================
%===================================================================================================================
\section{Problem of Time application: set of \K observables}
%===================================================================================================================
%===================================================================================================================

\noindent {\bf (Dirac) Observables} \cite{DiracObs} alias {\bf constants of the motion} alias {\bf perennials} \cite{Kuchar93, +Perennials, KucharObs} are any function(al)s of the canonical variables $\mD\lfloor \bQ, \bP \rfloor$ of the canonical variables (see footnote 1 of Paper I 
for this notation) such that, at the classical level, their Poisson brackets with all the constraint functions vanish (perhaps weakly \cite{I93}).  
For a theory with

\noindent total constraint set $\{{\cal C}_{\sfA}\}$, Dirac observables O = D($\bQ, \bP$) obey
\beq
\{ {\cal C}_{\sfA}, {\mO}\} = 0 \mbox{ } .  
\label{DirObs}
\eeq
Thus, for geometrodynamics  
\beq
\{{\cal H}(x), {\mO}\} =  0 \mbox{ } ,                   
\label{OH0}  
\eeq
\beq
\{{\cal H}_{\mu}(x), {\mO}\} = 0 \mbox{ } .                     
\label{OHi0}
\eeq
Justification of the name `constants of the motion' conventionally follows from the total Hamiltonian being
%
%In File R we want the total almost-Hamiltonian counterpart of this...
%
$H\lfloor \Lambda^{\sfA} \rfloor = \int_{\Sigma}\Lambda^{\sfA}{\cal C}_{\sfA}$, so that (\ref{DirObs}) implies 
\beq
\frac{d{\mO}}{dt}[Q(t), P(t)] = 0 \mbox{ } . 
\eeq

\mbox{ } 

\noindent{\bf Alternative Frozen Formalism Facet}. 
The operator-and-commutator counterparts of the above are then another manifestation of the Frozen Formalism Problem of classical canonical GR.  
[This is some sort of `Heisenberg' counterpart of the `Schr\"{o}dinger' Wheeler--DeWitt equation being frozen.]

\mbox{ }  

\noindent {\bf Kucha\v{r}'s Unicorn} I take this to be a sufficient set of Dirac observables/perennials to describe one's theory is termed.  
This follows from his quotation ``{\it Perennials in canonical gravity may have the same ontological status as unicorns 
-- a priori, these are possible animals, but a posteriori, they are not roaming on the earth}" \cite{Kuchar93}.  

\mbox{ }

\noindent Replace (\ref{DirObs}) with split conditions
\beq
\{{\cal Q}\muu\ma\md, \mO\} = 0 \mbox{ } ,
\label{Qobs}
\eeq
\beq
\{{\cal L}\mi\mn_{\sfZ}, \mO\} = 0  \mbox{ } . 
\label{Lobs}
\eeq
Usually there is but one quadratic constraint (per space point), though one could index it if needs be (see e.g. \cite{MRT}). 

\mbox{ } 

\noindent {\bf Kucha\v{r} observables} \cite{Kuchar93} are then as above except that only their brackets with the linear constraints (\ref{Lobs})  need vanish. 
 
\mbox{ }  

\noindent Kucha\v{r} then argued \cite{Kuchar93} for only the former needing to hold, in which case I denote the objects not by O = D$\lfloor \bQ, \bP\rfloor$ but by K$\lfloor \bQ, \bP \rfloor$, with the K standing for `Kucha\v{r} observable'.   
See also \cite{Kuchar93, KucharObs}.  

\noindent 
Beyond these arguments of sufficiency, I also use these in this series of papers as a technical half-way concept/construct in the formal 
and actual construction of Dirac observables.  

\mbox{ } 

\noindent As regards partial observables, one usually starts this discussion with {\bf true observables} (Rovelli 1991 \cite{Rov91}, see also \cite{Carlip90}) alias {\bf complete observables} (Rovelli 2002 \cite{Rov02}, and which at least Thiemann \cite{Thiemann} also calls evolving constant of the motion) classically involve operations on a system each of which produces a number that can be predicted if the state of the system is known.  
This conceptualization of observables is related to the above Dirac observables 
and should then be contrasted with the following much more cleanly distinct conception.

\mbox{ } 
 
\noindent Then {\bf Partial observables} themselves (Rovelli 1991 \cite{Rov91}) classically involve operation on the system that produces a number that is possibly totally unpredictable even if the state is perfectly known.  

\mbox{ }  

\noindent While the above definitions were more or less in place by 1991, the early 1990's and 2000's forms of the Problem of Time 
strategies that use these do themselves in part differ.
Since these approaches will largely not play a further role in the present article, I refer to \cite{Rov02,  Rovellibook, Dittrich, 
APOT}  for their further characterization and remaining difficulties.

Quantum-mechanically, each of the above two notions of observables carry over except that the entities whose predictabilities enter 
the definitions become quantum mechanical, the brackets become commutators and, in Rovelli's approach, the states are now taken to be specifically Heisenberg states.

I view this as a major first application of the understanding gained in Paper I about the shape variables and in Paper II about 
their conjugates to Problem of Time issues.  

\mbox{ } 

\noindent Now, shape variables and shape momenta have the additional interpretation as \K observables.
Then that the shape variables lucidly correspond to/are centred about geometrically significant configurations and their momenta lucidly correspond to changes of these acquires further significance.  One gets the sense that these are practically interesting observables and, 
at least sometimes, correspond to localized clusters.

Now, for the pure-shape case, (\ref{Qobs}) and (\ref{Lobs}) become 
\beq
\{{\ttH}, \mO\}       =  0 \mbox{ } ,
\label{OE0}  
\eeq
\beq
\{{\ttL}_{\mu}, \mO\} = 0 \mbox{ } ,                     
\label{OLi0}
\eeq
\beq
\{{\ttD}, \mO\} = 0 \mbox{ } .  
\label{OD0}
\eeq
Then \K observables O = K($\bQ, \bP$) solve (\ref{OLi0}) for the scaled case, and (\ref{OLi0}, \ref{OD0}) for the pure-shape case.
Dirac observables O = D($\bQ, \bP$) solve (\ref{OE0}, \ref{OLi0}) for the scaled case and (\ref{OE0}, \ref{OLi0}, \ref{OD0}) for the pure-shape case.  
This is because the Best Matching problem \cite{FileR} (the geometrodynamical case of which is the Thin Sandwich Problem, and which is a further facet of theProblem of Time) is solved for 1- and 2-$d$ RPM's, whether pure-shape or scaled, by \cite{FORD, Cones, FileR}.  
And that straightforwardly amounts to a construction and interpretation of a resolution of the problem of \K observables for that.  
This occurs in pure-shape RPM for precisely the set of all functions of the shape variables and the shape momenta, $\mK(\bS, \bP_{\sS})$.
Likewise, the set of  Kucha\v{r} observables for pure-shape RPM is precisely the set of all functions of the scale and shape variables and the 
scale and shape momenta, $\mK(\sigma, \bS, \mP_{\sigma}, \bP_{\sS})$.
The quantum counterpart of the above then `straightforwardly' involves some operator form for the canonical variables and commutators in place of 
Poisson brackets.

Note that here the best-matching problem is solved for 1- and 2-$d$ RPM's, whether pure-shape or scaled by results summarized in Paper I.  
And have been able to straightforwardly construct and interpret a resolution of the problem of observables in the sense of \K.  
The corresponding Kucha\v{r} observables are those quantities whose brackets with the linear constraints vanish.  
This occurs in pure-shape RPM for precisely the set of all functions of the shape variables and the shape momenta. 
Likewise, the set of  Kucha\v{r} observables for pure-shape RPM is precisely the set of all functions of the scale and shape variables and the 
scale and shape momenta.

I can spell out what all of these are for pure-shape and scaled RPM's in 1- and 2-$d$. 
The 1-$d$ pure-shape r-configuration spaces are \cite{06II} $\mathbb{S}^{N - 2}$ and suitable shape variables thereupon are the (ultra)spherical angles \cite{AF}, interpreted as functions of ratios of relative separations.  
This is as exemplified in Sec I.4 for 3- and 4-stop metroland cases.
The corresponding shape momenta are then as per Sec 2.

The 2-$d$ pure-shape r-configuration spaces are $\mathbb{CP}^{N - 2}$ and suitable shape variables fore these are the inhomogenous coordinates ${\mZ}^{\barr}$.  
To interpret these complex coordinates in terms of the $N$-a-gons, it is useful to pass to their polar forms,  
${\mZ}^{\barr} = {\cal R}^{\barr}\mbox{exp}(i\Phi^{\barr})$.  
Then the moduli are, again, ratios of relative separations, and the phases are now relative angles.  
In the specific case of the scalefree triangle, there is one of each, e.g. in coordinates based around the 1,23 clustering, these are 
\cite{TriCl} $\Theta = 2\,\mbox{arctan}(\rho_2/\rho_1)$ and $\mbox{arccos}\big( \brho_1 \cdot \brho_3 / \rho_1 \rho_3 \big)$.
The shape momenta for the $N$-a-gon are \cite{FileR} 
\beq
{\cal P}^{{\cal R}}_{\barp} = 
\left\{
\frac{\delta_{\barp\barq}}{1 + ||{\cal R}||^2}   - 
\frac{{\cal R}_{\barp}{\cal R}_{\barq}}{\{1 + ||{\cal R}||^2\}^2}
\right\}
{\cal R}_{\barq}^{*}
\mbox{ } \mbox{ } , \mbox{ } \mbox{ }  
{\cal P}^{\Theta}_{\widetilde{\sp}} = 
\left\{
\frac{\delta_{\overline{\sp}\overline{\sq}}}{1 + ||{\cal R}||^2} - 
\frac{{\cal R}_{\overline{\sp}}{\cal R}_{\overline{\sq}}}{\{1 + ||{\cal R}||^2\}^2}
\right\}
{\cal R}_{\overline{\sp}}{\cal R}_{\overline{\sq}}\Theta_{\widetilde{\sp}}^{*} \mbox{ } .  
\eeq

\noindent I gave the triangle in \cite{AHall} as a specific 2-d example; in the present seminar I give the quadrilateralland case 
as a larger and new specific example.  
This puts the program in \cite{Halliwell} into a whole-universe, nontrivially linearly constrained context.

\noindent The Gibbons--Pope version of this gets a fivefold interpretation in terms of quadrilaterals.   
All of these are intuitive and, for some configuration space regions, local, conditions.  
Namely, that of Sec I.5 for shapes and Sec 2 for momenta.
Then \K observables for this problem are functions of the form 
K($\chi, \beta, \phi, \psi, \pi_{\chi}, \pi_{\beta}, \pi_{\phi}, \pi_{\psi}$ alone).   

\noindent These all make for geometrically (in space) meaningful propositions and some are sometimes locally determinable/locally observable.  
\noindent Actual propositions involve approximate values of quantities, and this then rests on configuration space regions as studied in Paper I.

\noindent As regards the use of conserved quantities in preference to/alongside the momenta, 

\noindent 1) these, or functions thereof, commute also with the Hamiltonian constraint and are thus Dirac Observables.  
They manage to be this way via not encountering an obstruction from the potential term in \{O, H\}.    

\noindent 2) They feature in the kinematical quantization procedure, making them even more natural at the quantum level.  
For the sphere, these are the $SU(2)$ quantities ${\cal S}_{\sfA}$; for the quadrilateral, these are the $SU(3)$ quantities, especially the 
${\cal I}_{{\cal A}}$ and ${\cal Y}$ that remain conserved quantities for a wider range of potentials.
Here also e.g. for the sphere, $\Phi$ and $\Theta$ are not good operators, it is the unit Cartesian vectors that are.

\noindent A further issue here is what is the extent of overlap between kinematical quantization's \cite{I84} object selection and 
selection of observables.   
One's classical notion of observable is in each of the above cases to be replaced with the quantum one tied to a suitable commutation algebra in place of the classical Poisson algebra; this correspondence is however nontrivial (e.g. the two algebras may not be isomorphic) due to global considerations \cite{I84}.

%===================================================================================================================
%===================================================================================================================
\section{Equations of motion and conserved quantities for various potentials}
%===================================================================================================================
%===================================================================================================================

%===================================================================================================================
\subsection{$N$-stop metrolands and triangleland}
%===================================================================================================================

For triangleland [and suppressing (a)-labels]: $\Phi$-independence in the potential corresponds to there being no means for angular momentum to 
be exchanged between the subsystem composed of particles 2, 3 and that composed of particle 1.  
This corresponds to having an $SO(2)$ invariance (`special case')
If the potential is additionally $\Theta$-independent and so constant, one has the full $SO(3)$ invariance (`very special case') 
For $N$-stop metroland, there is likewise a sequence of special, very special, ... very$^{Y - 2}$ special potentials corresponding to 
$SO(2)$, $SO(3)$, ... $SO(Y)$ .
These above observations are useful as regards finding a nice range of analytic solutions of increasing complexity \cite{08I, 08II, 
+tri, AF, ScaleQM, 08III}. 
Moreover, e.g. for triangleland, there are in fact three particular $SO(2)$'s, corresponding to the three $\Phi$'s defined relative to the three DM axes present, albeit only one of these can be realized in any given model.
I will next consider the quadrilateralland counterparts of these statements.

%====================================================================================================
\subsection{Equations of motion for quadrilateralland in Gibbons--Pope type coordinates}
%====================================================================================================

The $\psi$-, $\phi$-, $\beta$- and $\chi$-equations are, respectively, 
\beq
\{\mbox{sin}^2\chi\,\mbox{cos}^2\chi\{\psi^* + \mbox{cos}\,\beta\,\phi^*\}/4\}^* = - {\pa \fV}/{\pa \psi}
\mbox{ } ,
\eeq
\beq
\{\mbox{sin}^2\chi\{\mbox{cos}^2\chi\{\phi^* + \mbox{cos}\,\beta\,\psi^*\} +  \mbox{sin}^2\chi\,\mbox{sin}^2\beta\,\phi^*\}/4\}^* 
= - {\pa \fV}/{\pa \phi}
\mbox{ } , 
\eeq
\beq
\{\mbox{sin}^2\chi\,\beta^*/4\}^* = 
\mbox{sin}^2\chi\,\mbox{sin}\,\beta\{\mbox{sin}^2\chi\mbox{cos}\,\beta\,\phi^* - \mbox{cos}^2\chi\,\psi^*\}\phi^*/4  - {\pa \fV}/{\pa \beta}
\mbox{ } , \mbox{ } \mbox{ and } 
\eeq
\beq
\chi^{**} = \mbox{sin}\,\chi\,\mbox{cos}\,\chi
\{\beta^{*\,2} + \mbox{cos}\,2\chi\,\{\phi^{*\,2} + \psi^{*\,2} + 2\,\phi^*\,\psi^*\mbox{cos}\,\beta\} + 2\mbox{sin}^2\chi\,\mbox{sin}^2\beta\,\phi^{*\,2} \}/4  - {\pa \fV}/{\pa \chi}
\mbox{ } .
\eeq
One of these can be supplanted by the energy first-integral,
\beq
{\chi^{* \, 2}}/{2} + {\mbox{sin}^2\chi}
\{\beta^{* \, 2} + \mbox{cos}^2\chi\{\phi^{*\,2} + \psi^{*\,2} + 2\,\phi^*\psi^*\mbox{cos}\,\beta\} + \mbox{sin}^2\chi\,\mbox{sin}^2\beta\,\phi^{*\, 2}\}\}/8 + \fV = \fE
\mbox{ } .
\eeq
(See \cite{AF} for the 4-stop metroland equations of motion, and \cite{08I} for the triangleland ones.)

%==================================================================================================================
\subsection{Which potentials realize which subgroups?}  
%==================================================================================================================

\noindent i)  For $\fV$ explicitly dependent on all of $\chi, \beta, \phi, \psi$, no isometry generator survives as a conserved quantity.  

\noindent ii) For $\fV$ $\psi$-independent, $\psi$ is a cyclic coordinate and yields one constant of the motion, 
\beq
\mbox{sin}^2\chi\,\mbox{cos}^2\chi\{\psi^* + \mbox{cos}\,\beta\,\phi^*\} =  C \mbox{ } .
\eeq
This corresponds to a $U(1)$ symmetry.  
I identify this $C$ as $2\,{\cal Y}$.

\noindent
iii) For $\fV$ $\phi$-independent, $\phi$ is a cyclic coordinate and yields another constant of the motion, 
\beq
\mbox{sin}^2\chi\{\mbox{cos}^2\chi\{\phi^* + \mbox{cos}\,\beta\,\psi^*\} +  \mbox{sin}^2\chi\,\mbox{sin}^2\beta\,\phi^*\} = K \mbox{ } .  
\eeq
This also corresponds to a $U(1)$ symmetry.  
I identify this constant $K$ as $4\,{\cal I}_3$.

\noindent 
iv) Potentials independent of both $\psi$ and $\phi$ yield both of these at once, corresponding to a $U(1) \times U(1)$ symmetry.  

\noindent 
Note that all of the symmetries considered so far may be viewed as phase factors in the complex representation, by which their $U(1)$ nature is rendered clear.

\noindent v) Potentials independent of both $\beta$ and $\phi$ yield three conserved quantities, corresponding to $SU(2)$ symmetry.  

\noindent vi) Potentials independent of all of $\beta$, $\phi$ and $\psi$ yield four, corresponding to $SU(2) \times U(1)$ symmetry.  

\noindent vii) If the potential is constant, one has all eight conserved quantities corresponding to the full $SU(3)$ isometry group. 

\noindent There are also $U$ and $V$ counterparts of all of the above, so that there are 3 versions of all the partial symmetries.  

%FFFFFFFFFFFFFFFFFFFFFFFFFFFFFFFFFFFFFFFFFFFFFFFFFFFFFFFFFFFFFFFFFFFFFFFFFFFFFFFFFFFFFFFFFFFFFFFFFFFFFFF
{            \begin{figure}[ht]
\centering
\includegraphics[width=0.55\textwidth]{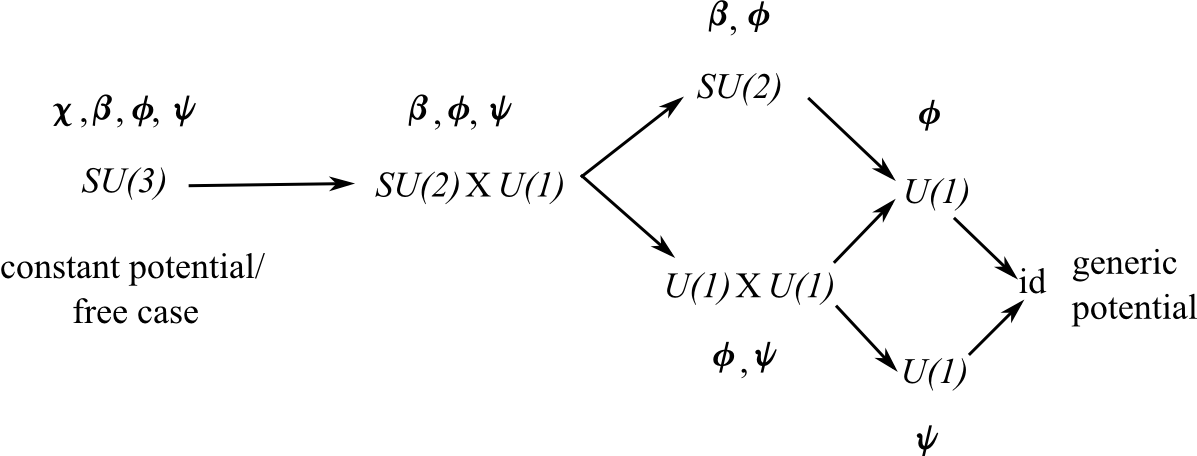}
\caption[Text der im Bilderverzeichnis auftaucht]{        \footnotesize{Flow diagram for the breakdown of the $SU(3)$ 
symmetry group due to various different potentials.  
The coordinates listed are those required to be absent from the potential. 
In fact, there are 3 different $SU(2) \times U(1)$ possibilities, each adapted to one of ${\cal I}$, ${\cal V}$ and ${\cal U}$.} }
\label{Fig1} \end{figure}          }
%FFFFFFFFFFFFFFFFFFFFFFFFFFFFFFFFFFFFFFFFFFFFFFFFFFFFFFFFFFFFFFFFFFFFFFFFFFFFFFFFFFFFFFFFFFFFFFFFFFFFFFFF

%==================================================================================================
\subsection{Hamiltonians with conserved quantities back-substituted in}
%==================================================================================================

For 3-stop metroland with $\varphi$-independent potential, $\fH = {\cal D}_{\sT\so\st} + \fV$, constant. 
For 4-stop metroland, with $\phi$-independent potentials, 

\noindent
\beq
\fH = {p_{\theta}^2}/{2} + {{\cal D}_{\phi}^2}/{2\mbox{sin}^2{\theta}} + \fV(\theta) \mbox{ } .
\eeq
For triangleland, with $\Phi$-independent potentials [$U(1)$-symmetric case] one has  \cite{TriCl, 08I}
\beq
\fH = {p_{\Theta}^2}/{2} + {{\cal J}^2}/{2\mbox{sin}^2{\Theta}} + \fV(\Theta) \mbox{ } .
\eeq
Each of these two has a constant case for the full $SO(3)$ symmetry. 
The question then is what is the quadrilateralland counterpart of these simplified (partly) symmetric cases.

One can now use the conserved quantity equations to write down Hamiltonians with more constants of the motion inside instead of momenta.

If there is a $U(1)$ symmetry of the $\psi$ type, 
\beq
\fH = \frac{p_{\chi}^2}{2} + \frac{2}{\mbox{sin}^2\chi}\left\{p_{\beta}^2 + \frac{1}{\mbox{sin}^2\beta}
\left\{p_{\phi}^2 + \frac{{\cal Y}^2}{4} - p_{\phi}{\cal Y}\,\mbox{cos}\,\beta\right\}\right\} + \frac{{\cal Y}^2}{2\mbox{cos}^2\chi}  + \fV(\chi, \beta, \phi) 
\mbox{ } , \mbox{ } \mbox{ ${\cal Y}$ constant } .
\eeq
If there is a $U(1)$ symmetry of the $\phi$ type, 
\beq
\fH = \frac{p_{\chi}^2}{2} + \frac{2}{\mbox{sin}^2\chi}\left\{p_{\beta}^2 + \frac{1}{\mbox{sin}^2\beta}
\{{\cal I}_{3}^2 + p_{\psi}^2 - 2 \, {\cal I}_{3}p_{\psi}\mbox{cos}\,\beta\}\right\} + \frac{2p_{\psi}}{2\mbox{cos}^2\chi}  + \fV(\chi, \beta, \psi) 
\mbox{ } , \mbox{ } \mbox{ ${\cal I}_3$ constant } .
\eeq
If there is a $U(1) \times U(1)$ symmetry, 
\beq
\fH = \frac{p_{\chi}^2}{2} + \frac{2}{\mbox{sin}^2\chi}\left\{p_{\beta}^2 + \frac{1}{\mbox{sin}^2\beta}
\left\{{\cal I}_{3}^2 + \frac{{\cal Y}_{\psi}^2}{4} - {\cal I}_{3}{\cal Y}\,\mbox{cos}\,\beta \right\}\right\} + \frac{{\cal Y}^2}{2\mbox{cos}^2\chi} 
+ \fV(\chi, \beta) \mbox{ } , \mbox{ } \mbox{ ${\cal I}_3$, ${\cal Y}$ constant } .
\eeq
If there is a $SU(2)$ symmetry, 
\beq
\fH = {p_{\chi}^2}/{2} + {2{\cal I}^2}/{\mbox{sin}^2\chi}  + {2p_{\psi}^2}/{\mbox{cos}^2\chi}  + \fV(\chi, \psi) 
\mbox{ } , \mbox{ } \mbox{ ${\cal I}$ constant } .
\eeq
If there is an $SU(2) \times U(1)$ symmetry, 
\beq
\fH = {p_{\chi}^2}/{2} + {2{\cal I}^2}/{\mbox{sin}^2\chi} + {{\cal Y}^2}/{2\mbox{cos}^2\chi}  + \fV(\chi) 
\mbox{ } , \mbox{ } \mbox{ ${\cal I}$, ${\cal Y}$ constant } . 
\eeq
The $SU(3)$ symmetry has $\fH$ constant.   

The simplest nontrivial case of $SU(2) \times U(1)$ symmetry can be straightforwardly represented as (via $u = \mbox{cos}^2\chi$)      
\beq
t^{\se\sm} - t^{\se\sm}_0 = - \frac{1}{2}\int \frac{\d u}{\sqrt{- {\cal Y}^2 + \{{\cal Y}^2 + 2\fW(u) - 4{\cal I}^2\}u - 2\fW(u) u^2}} \mbox{ } .
\label{uint}
\eeq

%============================================================================================================================================== 
%==============================================================================================================================================
\section{\underline{Key 10}: HO-type potentials} 
%==============================================================================================================================================
%==============================================================================================================================================

As explained in e.g. \cite{FileR} these are not HO's per se in the pure-shape case, since they have to be homogeneous of degree zero in order 
to be consistent.
This is attained by dividing the usual HO expression for the potential by the moment of inertia of the system (which subsequently turns out to be 
a constant).   
The scaled case has the usual HO potentials.  
\beq
\mbox{ } \mbox{ } \mbox{For 3-stop metroland \cite{FileR}, }
V = K_1 \mn_1^2/2 + K_2 \mn_2^2/2 + L \rho_1 \rho_2 = A + B\,\mbox{cos}\,2\varphi + C\mbox{sin}\,2\varphi  \hspace{3in}
\eeq
\beq
\mbox{for } \mbox{ } A = \{K_1 + K_2\}/2 \mbox{ } , \mbox{ } \mbox{ } B = \{K_2 - K_1\}/2 \mbox{ } , \mbox{ } \mbox{ } C = L/2 \mbox{ } .
\eeq
There is a special case with $SO(2) = U(1)$ symmetry, for $B = 0 = C$ i.e. $L = 0$ and $K_1 = K_2$, so that cluster 1 and cluster 2 have the same `constitution': the same Jacobi--Hooke coefficient per Jacobi cluster mass; here the `constituent springs'\!' potential contributions balance out to produce the constant potential, $\fV = A$. 
This is a kind of `homogeneity requirement' on the `structure' of the model universe.
For 4-stop metroland \cite{AF, ScaleQM},   
\beq
\fV = \sum\mbox{}_{a = 1}^3\{K_{a}\mn^{a\, 2}/2 + L_{a}\mn^{b}\mn^{c}\}
    = \ttA + \ttB\,\mbox{cos}\,2\theta + \ttC\,\mbox{sin}^2\,\theta\,\mbox{cos}\,2\phi + 
\ttD\,\mbox{sin}^2\theta\,\mbox{sin}\,2\phi + \ttE\,\mbox{sin}\,2\theta\,\mbox{cos}\,\phi + \ttF\,\mbox{sin}\,2\theta\,\mbox{sin}\,\phi 
\label{Bro}
\eeq 
\beq
\mbox{ for } \mbox{ } \ttA = \frac{1}{4}\left\{K_3 + \frac{K_1 + K_2}{2}\right\} \mbox{ } , \mbox{ } 
\ttB = \frac{1}{4}\left\{K_3 - \frac{K_1 + K_2}{2}\right\} \mbox{ } , \mbox{ } 
\ttC = \frac{K_1 - K_2}{4} \mbox{ } .  
\eeq
This has a special case with $U(1)$ symmetry, for $L_a = 0$ and $K_1 = K_2$ [i.e. $\ttC, \ttD, \ttE, \ttF = 0$],
$\fV = \ttA + \ttB\, \mbox{cos}\,2\theta$.  
It also has a very special case with $SO(3)$ symmetry, for $\ttB, \ttC, \ttD, \ttE, \ttF = 0$, i.e. $L_a = 0$ and $K_1 = K_2 = K_3$, for which 
high-symmetry situation the various potential contributions balance out to produce the constant, $\fV = \ttA$.  
\beq
\mbox{ } \mbox{ } \mbox{For triangleland \cite{+tri, 08III}, } \hspace{0.2in} 
\fV = K_1\mn_1\mbox{}^2/2 + K_2\mn_2\mbox{}^2/2 + L \underline{\mn}^1 \cdot \underline{\mn}^2 = 
      A + B\,\mbox{cos}\,\Theta + C\,\mbox{sin}\,\Theta\,\mbox{cos}\,\Phi \mbox{ } . \hspace{3in}
\eeq 
This has a special case with $U(1)$ symmetry, for $L = 0$ (i.e. $C = 0$), $\fV =  A + B\,\mbox{cos}\,\Theta$.  
It also has a very special case with $SO(3)$ symmetry, for $L = 0$, $K_1 = K_2$, i.e. $B, C = 0$, $\fV =  A$.  

\noindent Then for quadrilateralland, a parametrization of the HO-type potential at level of Jacobi vectors is
\beq
\fV =  \{K_1\mn_1^2 + K_2\mn_2^2 + K_3\mn_3^2\}/2 + L_1\bn_2 \cdot \bn_3 + L_2\bn_3 \cdot \bn_1 + L_3\bn_1 \cdot \bn_2 \mbox{ } .  
\eeq
I firstly note that Kuiper coordinates are {\sl very} HO-adapted:  
\beq
\fV = \{K_1N^1 + K_2N^2 + K_3N^3\}/2 + L_1\mbox{aniso}(23) + L_2\mbox{aniso}(31) + L_3\mbox{aniso}(12) \mbox{ } .  
\eeq
However, redundancy and non-adaptation of the kinetic term limit the usefulness of this expression.

The first 3 terms of this in Gibbons--Pope-type coordinates form the combination as the first three terms of (\ref{Bro}), since it involves solely 
the real parts for which the 2-$d$ 4 particle problem reduces to the 1-$d$ 4 particle one (in fact the mirror image identified version of this).
This rearrangement uses linear combinations of eqs (I.70-I.72).
For the other 3 cross-terms, however, the analysis has specific 2-$d$ character in contrast to (\ref{Bro})'s 1-$d$ character. 
These are, for H and K(M$^*$D) coordinates, and using eqs (I.73-I.75),
\beq
\fV =  
\left.
\left\{
L_3\mbox{sin}\,\beta\, \mbox{sin}^2\chi \mbox{cos}\,f_3 +  
L_1 \mbox{sin}\mbox{$\frac{\beta}{2}$} \mbox{sin}\,2\chi\, \mbox{cos}\,f_1 +  
L_2 \mbox{cos}\mbox{$\frac{\beta}{2}$} \mbox{sin}\,2\chi\, \mbox{cos}\,f_2
\right\}
\right/
2   \mbox{ }  .
\eeq
On the other hand, for K(T) coordinates, one has the $2  \leftrightarrow 3$ of the above, and likewise by Sec I.7's transpositions argument for the other choices of tree and of ratios.  
\noindent As regards HO-like potentials possessing particular symmetries, see Fig \ref{FlowD2}.
%
%FFFFFFFFFFFFFFFFFFFFFFFFFFFFFFFFFFFFFFFFFFFFFFFFFFFFFFFFFFFFFFFFFFFFFFFFFFFFFFFFFFFFFFFFFFFFFFFFFFFFFFFFFFFFFFFFFFFFFFFFFFFFFFFFFFFFFFFFFFFFF
{            \begin{figure}[ht]
\centering
\includegraphics[width=0.95\textwidth]{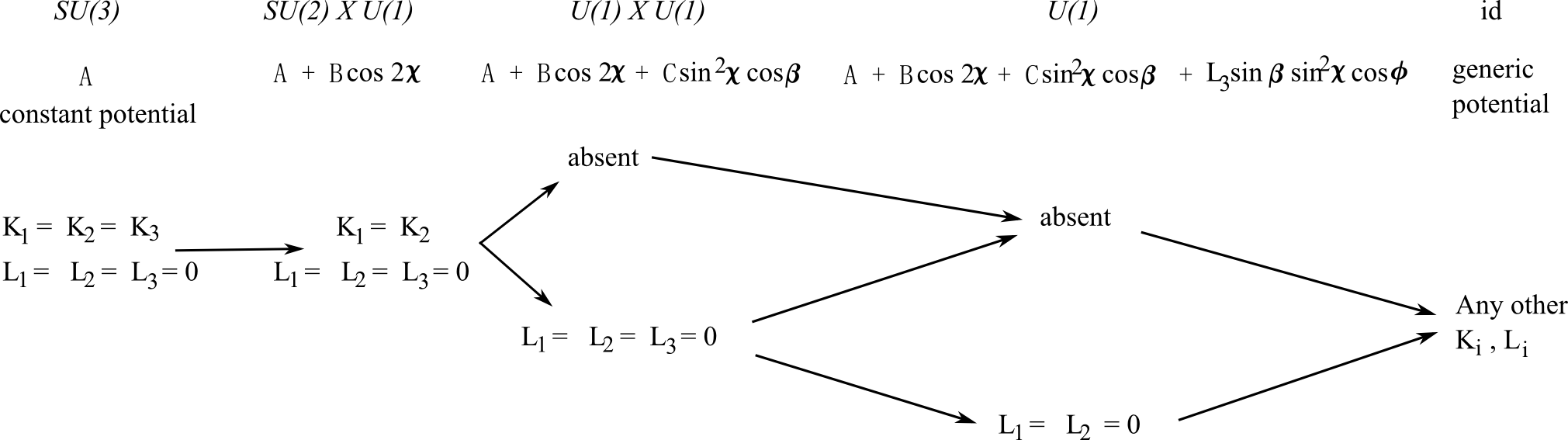}
\caption[Text der im Bilderverzeichnis auftaucht]{        \footnotesize{Flow diagram of the breakdown of the kinetic term's $SU(3)$ 
symmetry group due to various different HO potentials.
In the $SU(2) \times U(1)$ symmetric: if realized alongside at most the first term, it is the partial balance $K_1 = K_2$ of particular 
significance in the H-case as universe contents homogeneity, and in which case $\ttB = \{K_3 - K_1\}/4$. } }
\label{FlowD2} \end{figure}          }
%FFFFFFFFFFFFFFFFFFFFFFFFFFFFFFFFFFFFFFFFFFFFFFFFFFFFFFFFFFFFFFFFFFFFFFFFFFFFFFFFFFFFFFFFFFFFFFFFFFFFFFFFFFFFFFFFFFFFFFFFFFFFFFFFFFFFFFFFFFFFF

%===================================================================================================================
%===================================================================================================================
\section{Classical solutions for quadrilateralland}
%===================================================================================================================
%===================================================================================================================

%===================================================================================================================
\subsection{Geodesics of $\mathbb{CP}^{k}$ and their $N$-a-gonland interpretation}
%===================================================================================================================

For the general $\mathbb{CP}^k$ \cite{Warner82a}, geodesics through the origin are particularly simply expressed in complex form,  
\beq
Z^I = \tau C^I
\label{Warner}
\eeq
for $\tau$ a parameter and $C^I$ a constant vector.

%============================================================================================================================================
\subsection{Triangleland geodesics}
%============================================================================================================================================

(This is new to this program, insofar as \cite{08I} treated this as a real manifold.)    
Eliminating $\tau$ from (\ref{Warner}) in this case, and passing to the convenient spherical coordinates amounts to $\Phi$ being constant whilst $\Theta$ varies.   
The first of these conditions means that ${\cal J}$ = 0.  
These are the set of meridians with the origin being the D-pole corresponding to the underlying choice of clustering 
and the infinity being the M-pole antipodal to this.  
These are indeed a subset of the great circles that are well-known to be the geodesics in this case, and which 
were interpreted in terms of quadrilaterals in \cite{08I, +tri}.  
A particular such is the meridian of collinearity and another such is the meridian of isoscelesness. 
For later comparison, I furthermore note that these run between the clustering's two notions of collapse: 
$\rho_1 \longrightarrow 0$ and $\rho_2 \longrightarrow 0$.
I.e. from the arbitrarily sharp triangle to the arbitrarily flat one.  
The $N$-stop metroland spheres give (generalized) great circles but with complex formulations essentially absent (only present for $N$ = 3).

%=============================================================================================================================================
\subsection{\underline{Key 11}: Quadrilateralland geodesics}
%=============================================================================================================================================

\noindent In the quadrilateralland case, eliminating $\tau$ from (\ref{Warner}) and passing to the useful Gibbons--Pope coordinates 
amounts to $\psi$, $\phi$ and $\beta$ being constant whilst $\chi$ varies.  
The first three of these conditions imply that ${\cal Y}$ and ${\cal I}_{\sT\so\st}$ are both zero.
These also run between two collapsed cases, although now there is a diversity of such collapses available and of interpretations 
for these geodesics, according to the choices H or K and then of which common denominator to pick in making the subsequent two ratios 
(the usual five choices of Sec I.7).  
The 0 end is the 2 Jacobi distance collapse case and the $\infty$ end is the 1 Jacobi distance collapse complementary to it.  
Thus, these motions pick out the following.

\noindent For the usual H, this geodesic family corresponds to the posts to crossbar ratio increasing from 
`both posts collapsed to form a DD' of Fig I.8 o) to crossbar collapsed to form the rhombus of Fig I.8.g).  

\noindent For H(M$^*$D), this geodesic family corresponds to the crossbar and one post to the other post ratio increasing from
`one post and the crossbar collapsed to form M$^*$D' of e.g. Fig I.8.m) to the other post collapsing to form the triangle of Fig I.8.e).
(This one covers 2 cases at once, corresponding to symmetry of {\sl either} post collapsing).  

\noindent For the usual K, this geodesic family corresponds to the back and seat to leg ratio increasing from
`back and seat collapsed to form a T' of Fig I.8.q)  to        `leg-collapsed particle 3 onto T' triangle of Fig I.8.k).

\noindent For K(M$^{\sT}$D), this geodesic family corresponds to the back and leg to seat ratio increasing from
`back and leg collapsed to form a M$^{\sT}$ and a D' of Fig I.8.r) to `seat-collapsed T onto +' triangle of Fig I.8.l).

\noindent For K(M$^*$D), this geodesic family corresponds to the leg and seat to back ratio increasing from 
`leg and seat collapsed to form a M$^*$D' of Fig I.8.p)     to `back-collapsed DD' triangle of Fig I.8.j).

%==========================================================================================================================
\subsection{Time-traversal formulae}
%==========================================================================================================================

\noindent Some simple time-traversal cases that have analytical integrals (in terms of the emergent time) are as follows [from (\ref{uint}].

\noindent 1) For constant potential, ${\cal I}_{\sT\so\st\sa\sll}$ and ${\cal Y}$ are both 0, and then 
\beq
\chi = \sqrt{2W}\{t^{\se\sm} - t^{\se\sm}_0\} \mbox{ } .  
\eeq  
The shapes here are just $\chi$ changes as per above, and the time traversal confirms that these do not turn around, so the $\chi$ runs from one extreme value to the other.

\noindent 2) Formulae such as (precise trig/hyp functions involved depend on the signs of the various coefficients involved) 
\beq
\chi = \mbox{arccos}\big(\sqrt{W -  2{\cal I}_{\sT\so\st\sa\sll}}\mbox{sin}\big(-\sqrt{2W}\{ t^{\se\sm} - t^{\se\sm}_0 \}\big)\big)
\eeq 
for nonzero ${\cal I}_{\sT\so\st\sa\sll}$ and zero ${\cal Y}$ and still having constant potential.  

\noindent 3) Formulae such as 
\beq
\chi = \mbox{arccos}\big(\sqrt{1 -  2{\cal Y}^2/2W}\mbox{sin}\big(\sqrt{2W}\{ t^{\se\sm} - t^{\se\sm}_0 \}\big)\big)
\eeq
for the ${\cal I}_{\sT\so\st\sa\sll} \longrightarrow {\cal Y}$ of the above.
\beq 
4) \hspace{1in} \chi = \mbox{arccos}
\left(
\sqrt{\frac{\{\mbox{exp}(\sqrt{C}\{ t^{\se\sm} - t^{\se\sm}_0 ) - B \}\{ \mbox{exp}(\sqrt{C}\{ t^{\se\sm} - t^{\se\sm}_0 ) - B - 2A\sqrt{C}   \}}{2\sqrt{C}\{B + 2 \sqrt{C}\}}}
\right)   \hspace{3in} \mbox{ }  
\eeq
for both of these conserved quantities being nonzero, and where $A = - {\cal Y}^2$, $B = 2\fW - 4{\cal I}^2 + {\cal Y}^2$ and $C = - 2\fW$.
4-stop metroland and triangleland are analogous to zero extra charge cases of the above.

%=====================================================================================================================================
\subsection{Further solutions}
%=====================================================================================================================================

\noindent Unlike for 4-stop metroland and triangleland, where there is a range of further classically-tractable solutions \cite{AF, 08I, FileR}, 
further cases for quadrilateralland, and the simplest HO counterparts, give at best combinations including elliptic functions (checked with Maple).  
The QM of these isn't analytically tractable either.  
However, the direction I will take is treating the HO potentials as small perturbations about the free case at the quantum level.

The following Figure provides useful qualitative analysis of ${\cal Y}$ = 0 and $\neq 0$ and ${\cal I}_{\sT\so\st\sa\sll} = 0$ and 
$\neq 0$ (\underline{Key 12}).  

%FFFFFFFFFFFFFFFFFFFFFFFFFFFFFFFFFFFFFFFFFFFFFFFFFFFFFFFFFFFFFFFFFFFFFFFFFFFFFFFFFFFFFFFFFFFFFFFFFFFFFFF
{            \begin{figure}[ht]
\centering
\includegraphics[width=1.1\textwidth]{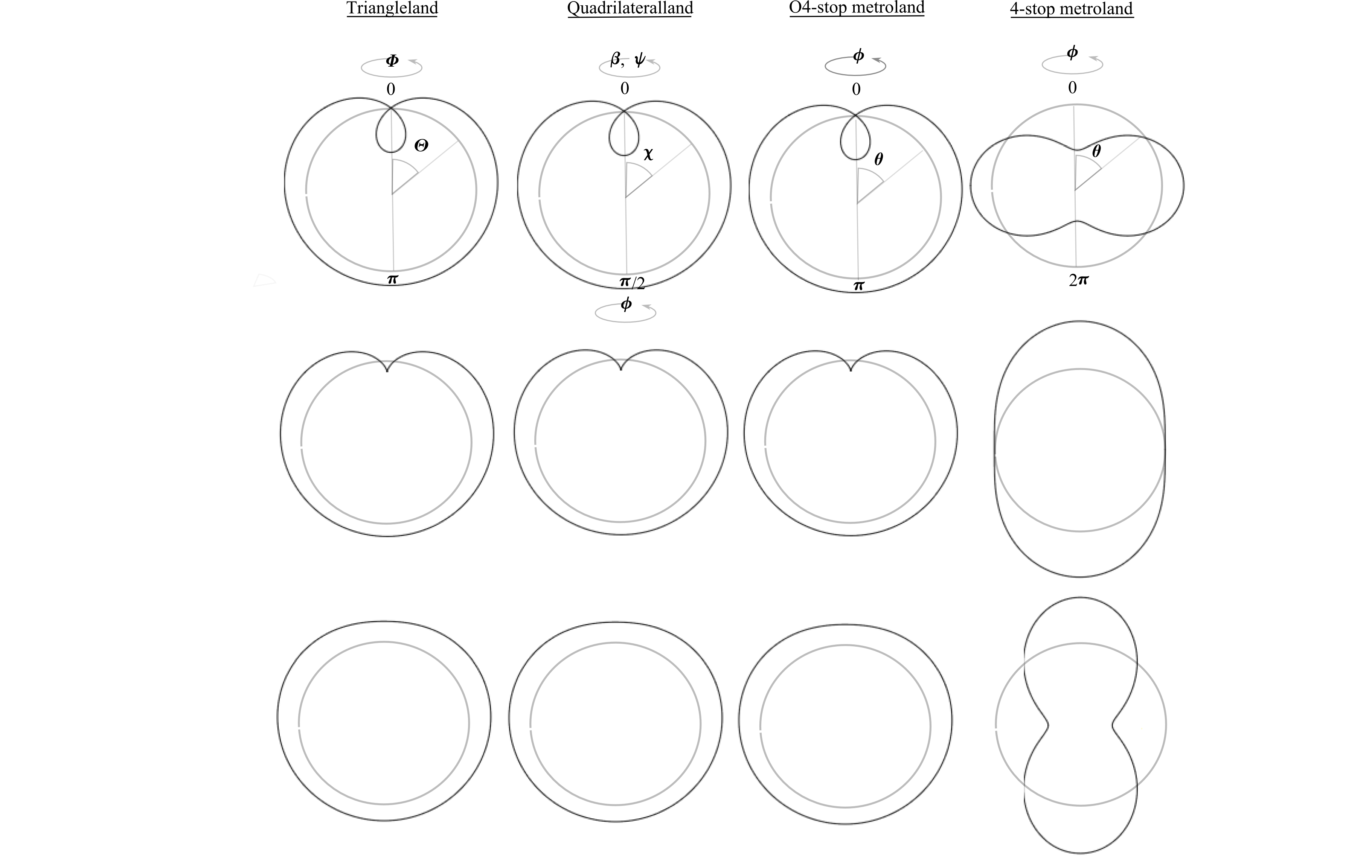}
\caption[Text der im Bilderverzeichnis auftaucht]{        \footnotesize{Qualitative behaviour of some of the simpler HO dynamics: 
for the potential $\ttA + \ttB\,\mbox{cos}\, 2\chi$ for various ratios of $\ttA$ and $\ttB$.  
The Quadrilateralland case of this (column 2) is usefully compared with the corresponding potentials from two existing works and a further variant.  
Namely, triangleland (column 1 \cite{+tri}), 4-stop metroland (column 4, \cite{AF}) and O4-stop metroland (column 3, also new to this 
Paper).  
Furthermore, a centrifugal spike as felt for $L_{\sT\so\st} \neq 0$ in ordinary mechanics or spherical mechanics, ${\cal S}_{\sT\so\st} \neq 0$ triangleland and ${\cal D}_{\sT\so\st} \neq 0$ for 4-stop metroland.  
That the on-sphere cases of this have a spike at {\sl each} pole is standard.
This occurs for the triangleland case and the 4-stop metroland case.  
The O4-stop metroland case has just the North Pole spike.  
A feature that first occurs in this paper's quadrilateralland case is having {\sl independent} spikes at each `pole'.  
Namely, a spike felt for ${\cal I}_{\sT\so\st\sa\sll} \neq 0$ at the `North pole' $\chi = 0$ and a different spike felt by ${\cal Y} \neq 0$ 
at the `South pole' $\chi = \pi/2$.  
I note that the O4-stop metroland case is recovered in the study of the collinear submanifold of the quadrilateralland case.
For contrast, I also note that 4-stop metroland itself is the odd one out in having peanut/egg/tyre shaped potentials, with the tyre 
alone possessing two wells, whereas all the other cases are heart/egg shaped potentials with at most one well.

In this paper's quadrilateralland case, the motions are qualitatively as follows. 
In the cross-section for conserved quantities being 0, and `round and round' in the other cases.  
That the $SU(2)$ round and round is some higher-$d$ $\mathbb{S}^2$ analogue of axisymmetry. 
The $U(1)$ of the $\psi$ is standard axisymmetry. 
\noindent [Note: each of these corresponds to going round-and-round in {\sl different} suppressed dimensions.] } }
\label{Temporary} \end{figure}          }
%FFFFFFFFFFFFFFFFFFFFFFFFFFFFFFFFFFFFFFFFFFFFFFFFFFFFFFFFFFFFFFFFFFFFFFFFFFFFFFFFFFFFFFFFFFFFFFFFFFFFFFFFFFFFFFFFFFF

%===================================================================================================================
%===================================================================================================================
\section{Conclusion}
%===================================================================================================================
%===================================================================================================================

%===================================================================================================================
\subsection{Contents summary}
%===================================================================================================================

The shape momenta conjugate to the shape variables studied in Paper I were considered.  
Physically, these are relative dilational momenta, relative angular momenta and mixtures of these.  
Conserved quantities for RPM's in 1-$d$ were also considered; these are particular combinations of the preceding.  
Functions of the shape variables and shape momenta can be used to resolve the Problem of Observables in the sense of \K for these models.  
This paper consolidates the above for small RPM's and is the first place to extend as far as the explicit quadrilateralland examples of these things.  
This is a substantial extension due to how quadrilateralland is the smallest RPM to simultaneously possess nontrivial subsystem structure and linear constraints in the quantum cosmologically significant case. 
It also has more general and typical mathematics for an $N$-a-gonland than triangleland does (since its shape space is $\mathbb{CP}^2$ whilst 
triangleland's is $\mathbb{CP}^1$, which is atypical through also being $\mathbb{S}^2$).

Quadrilateralland's isometry group is then (\underline{Key 9}), more or less, $SU(3)$; more precisely, it is $SU(3)/\mathbb{Z}_3$.    
Thus quadrilateralland exhibits a number of parallels with the Particle Physics of the strong force, 
though I use the more descriptive name `extra charge' instead of hypercharge.
One new question addressed in this paper then is how to interpret the $SU(3)$ quantities in terms of the quadrilateral.   
Gibbons--Pope-type coordinates are useful in this investigation - they have clear geometrical interpretation in quadrilateralland 
terms, and they are additionally cyclic coordinates for some of the simpler potentials
I gave the geometric interpretations of the momenta and conserved quantities in terms of these, for Jacobi H coordinates and Jacobi K coordinates 
for the various ratio choices, and then what various of the $SU(2)$ quantities are in terms of these.  
The quantum form of ${\cal I}_{\sT\so\st\sa\sll}$ that I provide in this paper is useful in building Paper III's time-independent 
Schr\"{o}dinger equation.

I considered the multi HO-like potential for quadrilateralland in terms of the useful intrinsic Gibbons--Pope coordinates (\underline{Key 10}).  
\noindent I interpreted a geodesics result for $\mathbb{CP}^2$ (\underline{Key 11}) in terms of quadrilaterals, and link it to the various collapses of the quadrilateralland trees.
\noindent In the qualitative dynamics figure  \ref{Temporary} for the simpler HO-like potential (\underline{Key 12}), I provide a nice extension of work in \cite{AF, +tri} with one extra case necessitated so as to parallel the quadrilateral better: the qualitative analysis of 
HO-type potentials on O4-stop metroland (i.e. the mirror-image-identified version).  
The new feature in quadrilateralland is two repulsive spikes that are felt independently of each other according to whether the 
system in question has each of the two types of charge: angular charge, paralleling isospin, and extra angular charge, paralleling hypercharge.  
\noindent These qualitative dynamics figures also combine nicely with the tessellation interpretations of the 4-stop metroland and triangleland 
shape spheres and of the characterized submanifolds of $\mathbb{CP}^2$ as provided in Paper I.

%=====================================================================================================================
\subsection{Outline of further applications of this paper's quadrilateralland work}  
%=====================================================================================================================

\noindent The $N$-a-gonland interpretation of conserved quantities and of corresponding quantum numbers remains open.  
\noindent For higher-$N$ cases, $SU(N)$ mastery including first parts of \cite{MF03b} but there is now no ready analogue of Gibbons--Pope 
coordinates available... 
\noindent The geodesics result used \cite{Warner82a} is general-$N$. 
\noindent Some further results on the classical dynamics on $\mathbb{CP}^N$ can be found in e.g. \cite{CPNDyn}. 

%\mbox{ } 

\noindent QM needs conserved quantity study and classical solutions as back-up.
Some problem of time strategies require classical solutions (semiclassical approach) or work at the classical level 
(internal time), some observables approaches.

\noindent The Schr\"{o}dinger equation for quadrilateralland is treated in Paper III.  
The equation itself is \underline{Key 13}, and is built from this paper's expression for ${\cal I}_{\sT\so\st\sa\sll}$ and using conformal invariance.  
This is then solvable in terms of Jacobi polynomials and Wigner D-functions for the free case (\underline{Key 14}), and a perturbative 
scheme can be set up within this sort of methods of mathematical physics so as to study small multi-HO like potentials (\underline{Key 15}).  
These are quite clearly extensions of importance of the Keys provided in the present seminar.  

\mbox{ } 

\noindent Papers I to III are then useful for subsequent investigations of Problem of Time in Quantum Gravity strategies and various other 
quantum-cosmological issues.  
These are mostly in Paper IV, with a bit about regions and uniformity in this paper, \K observables in Paper II and the \NSI and peakedness in Paper III.  
Particular such application for Quadrilateralland and $N$-a-gonland are to timeless approaches to the Problem of Time in Quantum Gravity --  
semiclassical, histories, records, observables and combined approaches (including Halliwell's \cite{Halliwell, AHall}), qualitative models of structure formation in Quantum Cosmology, and to the robustness study based on the \{$N$ -- 1\}-a-gon model lying inside the $N$-a-gon one.   
I note that complex projective mathematics (the present seminar involving the simplest RPM model paper with nontrivial such) will underlie this robustness study. 

\mbox{ }

\noindent {\bf Acknowledgements}: I thank those close to me for 
being supportive of me whilst this work was done.   
Eduardo Serna for discussions and for contributing to some of the calculations and checking other of the calculations. 
Professors Don Page and Gary Gibbons for material about $\mathbb{CP}^2$ in 2005 and 2007.
Dr Julian Barbour for introduction to RPM´s in 2001.  
Professors Enrique Alvarez and Marc Lachi\`{e}ze-Rey for discussions. 
Anya Ermakova for recommending, and presenting a copy of, \cite{Kvothe}.
Professors Belen Gavela, Malcolm MacCallum, Don Page, Reza Tavakol, Dr Jeremy Butterfield and especially Marc Lachi$\grave{\me}$ze-Rey for support with my career.  
This work was funded by a grant from the Foundational Questions Institute (FQXi) Fund, a donor-advised fund of the Silicon Valley Community Foundation on the basis of proposal FQXi-RFP3-1101 to the FQXi.  
I thank also Theiss Research and the CNRS for administering this grant, and Professors Marc Lachi$\grave{\me}$ze-Rey and David Langlois for APC travel money used for part of this project.

%=====================================================BIBLIOGRAPHY==========================================================================


\begin{thebibliography}{99}
%===========================================================================================================================================

\footnotesize

\bibitem{QuadI}              E. Anderson, Paper I.

\bibitem{RPMConcat}          J.B. Barbour and B. Bertotti, Proc. Roy. Soc. Lond. {\bf A382} 295 (1982);  
%
                             J.B. Barbour, Class. Quantum Grav. {\bf 11} 2853 (1994);
%
                             C. Kiefer, {\it Quantum Gravity} (Clarendon, Oxford 2004);  
%
                             E. Anderson, Class. Quantum Grav. {\bf 23} (2006) 2469, gr-qc/0511068; 
%
                             {\bf 24} 2935 (2007), gr-qc/0611007;
%
                             {\bf 26} 085015 (2009), arXiv:0810.4152;  
%
                             {\bf 27} 045002 (2010), arXiv:0905.3357;
%
                             {\bf 28} 185008 (2011), arXiv:1101.4916;
%
                             Int. J. Mod. Phys. {\bf D18} 635 (2009), arXiv:0709.1892;   
%
                             in {\it Proceedings of the Second Conference on Time and 
                             Matter}, ed. M. O'Loughlin, S. Stani\v{c} and D. Veberi\v{c} 
                             (University of Nova Gorica Press, Nova Gorica, Slovenia 2008), arXiv:0711.3174;  
%
                             for Proceedings of Paris 2009 Marcel Grossman Meeting, in Press, arXiv:0908.1983;
%
                             arXiv:1009.2161; 
% 
                             arXiv:1102.2862;    
%
                             arXiv:1205.1256;
%
                             arXiv:1209.1266;  
%
                             S.B. Gryb, arXiv:0804.2900; 
%
                             J.B. Barbour and B.Z. Foster, arXiv:0808.1223.  
%
                             J. Barbour, arXiv:1105.0183.

\bibitem{B03}                J.B. Barbour, Class. Quantum Grav. \textbf{20} 1543 (2003), gr-qc/0211021. 

\bibitem{06II}                E. Anderson  Class. Quantum Grav. {\bf 23} 2491 (2006), gr-qc/0511069.    

\bibitem{TriCl}               E. Anderson, Class. Quantum Grav. {\bf 24} 5317 (2007), gr-qc/0702083. 

\bibitem{FORD}               E. Anderson, Class. Quantum Grav. {\bf 25} 025003 (2008), arXiv:0706.3934.

\bibitem{08I}                E. Anderson, Class. Quantum Grav. {\bf 26} 135020 (2009), arXiv:0809.1168.   

\bibitem{AF}                 E. Anderson and A. Franzen, Class. Quantum Grav. {\bf 27} 045009 (2010), arXiv:0909.2436. 

\bibitem{+tri}               E. Anderson, Gen. Rel. Grav. {\bf 43} 1529 (2011), arXiv:0909.2439.  

\bibitem{Cones}              E. Anderson, arXiv:1001.1112.

\bibitem{ScaleQM}            E. Anderson, Class. Quantum. Grav. {\bf 28} 065011 (2011), arXiv:1003.4034.

\bibitem{08II}               E. Anderson, Class. Quantum Grav. {\bf 26} 135021 (2009) gr-qc/0809.3523.

\bibitem{08III}              E. Anderson, arXiv:1005.2507. 

\bibitem{SemiclIII}          E. Anderson, Class. Quantum Grav. {\bf 28} 185008 (2011),  arXiv:1101.4916. 

\bibitem{FileR}              E. Anderson, arXiv:1111.1472.

\bibitem{AHall}              E. Anderson, Class. Quant. Grav., {\bf 29} 235015 (2012), arXiv:1204.2868;   
%
                             Invited seminar at the `Do we need a Physics of Passage' Conference at Cape Town, December 2012, arXiv:1306.5816.
  
%================================================================================================================================================

\bibitem{Smith60}            F.T. Smith, Phys. Rev. {\bf 120} 1058 (1960). 

\bibitem{MacFarlane}         A.J. MacFarlane, J. Phys. A: Math. Gen. {\bf 36} 7049 (2003).  

\bibitem{PS}                 M.E. Peskin and D.V. Schroeder, {\it An Introduction to Quantum Field Theory} 
                            (Perseus Books, Reading, Massachusetts  1995).   

\bibitem{GiPo}               G.W. Gibbons and C.N. Pope, Commun. Math. Phys. {\bf 61} 239 (1978); 
%
                             C.N. Pope, Phys. Lett. {\bf 97B}  417 (1980).  

\bibitem{RWR}                J.B. Barbour, B.Z. Foster and N. \'{O} Murchadha, Class. Quantum Grav. 
                             {\bf 19} 3217 (2002), gr-qc/0012089; 
%
                             E. Anderson, Gen. Rel. Grav. {\bf 36} 255, gr-qc/0205118; 
%
                             Phys. Rev. {\bf D68} 104001 (2003),  gr-qc/0302035;
%
                             in {\it General Relativity Research Trends, Horizons in World 
                             Physics} {\bf 249} ed. A. Reimer (Nova, New York 2005), gr-qc/0405022;
%
                             Stud. Hist. Phil. Mod. Phys. {\bf 38} 15 (2007), gr-qc/0511070;
%
                             in ``Classical and Quantum Gravity Research", ed. M.N. 
                             Christiansen and T.K. Rasmussen (Nova, New York 2008), arXiv:0711.0285; 
%                             
                             E. Anderson, J.B. Barbour, B.Z. Foster and N. \'{O} Murchadha, Class. Quantum Grav. {\bf 20} 157 (2003), 
                              gr-qc/0211022;  
%
                             E. Anderson, J.B. Barbour, B.Z. Foster, B. Kelleher and N. \'{O} Murchadha, 
                             Class. Quantum Grav {\bf 22} 1795 (2005), gr-qc/0407104; 
%   
                             E. Anderson, ``A Note on Variational Methods Underpinning Shape Dynamics", forthcoming; 
%
                             J.B. Barbour and N. \'{O} Murchadha, arXiv:1009.3559.  

\bibitem{Kuchar93}           K.V. Kucha\v{r} 1993, in {\it General Relativity and Gravitation 1992},  
                             ed. R.J. Gleiser, C.N. Kozameh and O.M. Moreschi M (Institute of Physics Publishing, Bristol 1993), gr-qc/9304012.

\bibitem{K92}                K.V. Kucha\v{r}, in {\it Proceedings of the 4th Canadian Conference on 
                             General Relativity and Relativistic Astrophysics} ed. G. Kunstatter, D. 
                             Vincent and J. Williams (World Scientific, Singapore 1992). 

\bibitem{I93}                C.J. Isham, in {\it Integrable Systems, Quantum Groups and Quantum Field Theories}  
                             ed. L.A. Ibort and M.A. Rodr\'{\i}guez (Kluwer, Dordrecht 1993), gr-qc/9210011.

\bibitem{APOT}                E. Anderson, in {\it Classical and Quantum Gravity: Theory, Analysis and Applications}  
                             ed. V.R. Frignanni (Nova, New York 2011),arXiv:1009.2157; 
%                             
                             Invited Review in Annalen der Physik, {\bf 524} 757 (2012),  arXiv:1206.2403.

\bibitem{QuadIII}            E. Anderson and S.A.R. Kneller, arXiv:1303.5645.

\bibitem{QuadIV}             E. Anderson, forthcoming. 

\bibitem{Dragt}              A.J. Dragt, J. Math. Phys. {\bf 6} 533 (1965). 

\bibitem{WheelerInt}          J. Bi\v{c}\'{a}k, `` The art of science: interview with Professor John Archibald Wheeler", 
                              Gen. Rel. Grav. {\bf 41} 679 (2009), arXiv:1105.4532.
   
\bibitem{Kvothe}              P. Rothfuss, {\it The Name of the Wind} (Orion, London 2007).  

\bibitem{Macfarlane68}        A.J. MacFarlane, Comm. Math. Phys. {\bf 11} 91 (1968).

\bibitem{MF79}               A.J. MacFarlane, Nu. Phys. {\bf B152} 145 (1979).  

%============================================== On observables

\bibitem{DiracObs}           P.A.M. Dirac,  Rev. Mod. Phys. {\bf 21} 392 (1949).  

\bibitem{+Perennials}         P. H\'{a}j\'{\i}\v{c}ek, J. Math. Phys. {\bf 36} 4612 (1996), gr-qc/9412047;
%
                              C.J. Isham and P. H\'{a}j\'{\i}\v{c}ek, J. Math. Phys. {\bf 37} 3522 (1996), gr-qc/9510034;  
%   
                              J. Pullin and R. Gambini, in {\it 100 Years of Relativity.  
                              Space-Time Structure: Einstein and Beyond} ed. A Ashtekar (World Scientific, Singapore 2005).  

\bibitem{KucharObs}           K.V. \K, J. Math. Phys. {\bf 22} 2640 (1981); 
%
                              C.G. Torre, Phys. Rev. {\bf D48} 2373 (1993), gr-qc/9306030; 
%
                              Phys. Rev. Lett {\bf 70} 3525 (1993);  
%
                              C.G. Torre and I.M. Anderson, Commun. Math. Phys. {\bf 176} 479 (1996), gr-qc/9404030;    
%
                              P. H\'{a}j\'{\i}\v{c}ek, gr-qc/9903089;
%
                              P. H\'{a}j\'{\i}\v{c}ek and J. Kijowski, Phys. Rev. {\bf D61} 024037 (2000), gr-qc/99080451;  
%
                              G. Belot and J. Earman, in {\it Physics Meets Philosophy at the 
                              Planck Scale}, ed. C. Callender and N. Huggett, Cambridge University Press (2000).
%                                                           
                              S. Carlip, Rept. Prog. Phys. {\bf 64} 885 (2001), gr-qc/0108040. 
%
                              J. Earman, Phil. Sci. {\bf 69} S209 (2002);
%                             
                              H. Farajollahi, gr-qc/0406024;  
% 
                              C. W\"{u}threich, ``Approaching the Planck Scale from a Generally Relativistic Point of View: A Philosophical 
                              Appraisal of Loop Quantum Gravity" (Ph.D Thesis,  Pittsburgh 2006);      
%
                              J.B. Barbour and B.Z. Foster, arXiv:0808.1223.  

\bibitem{MRT}                 M. Montesinos, C. Rovelli and T. Thiemann, Phys. Rev. {\bf D60} 044009 (1999), gr-qc/9901073.

\bibitem{Rov91}               C. Rovelli, p. 126 in {\it Conceptual Problems of Quantum 
                              Gravity} ed. A. Ashtekar and J. Stachel (Birkh\"{a}user, Boston, 1991);   
%
                               Phys. Rev. {\bf D43} 442 (1991); 
%
                              {\bf D44} 1339 (1991).

\bibitem{Carlip90}            S. Carlip Phys. Rev. {\bf D42} 2647 (1990); 
% 
                              Class. Quantum Grav. {\bf 8} 5 (1991).   
                       
\bibitem{Rov02}               C. Rovelli, Phys. Rev. {\bf D65} 044017 (2002), arXiv:gr-qc/0110003; 
%
                              {\bf 65} 124013, gr-qc/0110035.                     

\bibitem{Thiemann}            T. Thiemann, {\it Modern Canonical Quantum General Relativity} (Cambridge University Press, Cambridge 2007). 

\bibitem{Rovellibook}        C. Rovelli, {\it Quantum Gravity} (Cambridge University Press, Cambridge 2004).    

\bibitem{Dittrich}            B. Dittrich, Class. Quant. Grav. {\bf 23} 6155 (2006), gr-qc/0507106.

%======================================== Some technical and conceptual back-up... 

\bibitem{Halliwell}           J.J. Halliwell and J. Thorwart, Phys. Rev. {\bf D65} 104009 (2002), gr-qc/0201070;     
%
                              J.J. Halliwell, in {\it The Future of Theoretical Physics and Cosmology} 
                              (Stephen Hawking 60th Birthday Festschrift volume) ed. G.W. Gibbons, E.P.S. 
                              Shellard and S.J. Rankin (Cambridge University Press, Cambridge 2003); 
%
                              Phys. Rev. {\bf D80} 124032 (2009), arXiv:0909.2597; 
%
                              J. Phys. Conf. Ser. {\bf 306} 012023 (2011), arXiv:1108.5991.

\bibitem{I84}                 C.J. Isham, in {\it Relativity, Groups and Topology {II}} 
                              ed. B. DeWitt and R. Stora (North-Holland, Amsterdam 1984).  

\bibitem{Warner82a}           N.P. Warner, Proc. Roy. Soc. Lond. {\bf 1383} 207 (1982).  
 
\bibitem{MF03b}               A.J. MacFarlane, J. Phys. A: Math. Gen. {\bf 36} 9689 (2003).  

\bibitem{CPNDyn}             See e.g. P. Foth, J. Math. Phys. {\bf 43} 3124 (2002); 
%
                             J.M. Isidro, Phys.Lett. {\bf A317} (2003) 343 quant-ph/0307172; 
%
                             www-fourier.ujf-grenoble.fr/$\widetilde{\mbox{ }}$faure/articles/acta$\underline{\mbox{ }}$02.ps.gz.

      
\end{thebibliography}
\end{document}